\documentclass[12pt]{amsart}
\usepackage{mathrsfs}
\usepackage{epsfig}
\begin{document}
 \topmargin=5mm
 \oddsidemargin=6mm
 \evensidemargin=6mm
\baselineskip=26pt
\parindent 20pt
{\flushleft{\Large\bf  Generalized Super Bell Polynomials with Applications to Superymmetric Equations  }}\\[12pt]
{\large\bf  Engui Fan\footnote{\ \ Corresponding author and E-mail
address: \  faneg@fudan.edu.cn} }\\[8pt]
{\small  School of Mathematical Sciences, Institute of Mathematics
and Key Laboratory of Mathematics for Nonlinear Science, Fudan
University, Shanghai,
200433, P.R.  China} \\[8pt]
{\large\bf  Y. C. Hon }\\[4pt]
{\small Department of Mathematics, City University of Hong Kong,
Hong Kong, P.R. China}\vspace{6mm}
\begin{center}
\begin{minipage}{5.2in}
\baselineskip=20pt { \small {\bf Abstract.}   In this paper, we
introduce  a class of new generalized super Bell polynomials on a
superspace, explore  their properties, and show that they are a
natural and effective tool to systematically investigate
integrability of supersymmetric equations. The connections between
the super Bell polynomials and super bilinear representation,
bilinear B\"{a}cklund transformation, Lax pair and infinite
conservation laws of supersymmetric equations are established.  We
take supersymmetric KdV equation and supersymmetric
sine-Gordon equation to illustrate  this  procedure.  }
\end{minipage}\\[24pt]
\end{center}
%%%%%%%%%%%%%%%%%%%%%%%%%%%%%%%%%%%%%%%%%%%%%%%%%%%%%%%%%%%%%%%%%%%%%%%%%%%%%%%%%%%%%%%%%%%%%%%%%%%%
{\bf\large 1. Introduction}\\

The supersymmetry  represents  a kind of symmetrical characteristic
between boson and fermion in physics.  The concept of supersymmetry
was originally introduced and developed for applications in
elementary particle physics thirty years ago
\cite{Ramond}--\cite{Wess}.  It is found that supersymmetry can be
applied to a variety of problems such as relativistic,
non-relativistic physics and nuclear physics. In recent years,
supersymmetry has been a subject of considerable interest both in
physics and mathematics.  The mathematical formulation of the
supersymmetry is based on the introduction of Grassmann variables
along with the standard ones \cite{Berezin}. In a such way, a number
of well known mathematical physical equations have been generalized
into the supersymmetric analogues, such as supersymmetric versions
of sine-Gordon, KdV, KP hierarchy, Boussinesq, MKdV  etc.   It has
been shown that these supersymmetric integrable systems possess
bi-Hamiltonian structure, Painlev\'{e} property, infinite many
symmetries, Darboux transformation, B\"{a}cklund transformation,
bilinear form,  super soliton solutions and super quasi-periodic
solutions \cite{Manin}--\cite{Liu4}.   In our present paper,  we
investigate  the integrability of supersymmetric equations by using
a class of super Bell polynomials which are a multidimensional and
super generalization of  ordinary Bell polynomials.

The ordinary Bell polynomials introduced by Bell during the early
1930s  are  a class of exponential polynomials, which are specified
by a generating  function and exhibit important properties
\cite{Bell}.   The Bell polynomials  have
  been exploited  in combinatorics, statistics and other fields
  \cite{Abr}-\cite{Com}.  Some generalized forms of Bell polynomials already appeared in literature
  \cite{Kolbig}-\cite{Wang}.
 More recently  Lambert, Gilson et al found that  the  Bell polynomials
  also play important role in  the characterization of bilinearizable equations.
  They presented  an alternative
procedure based on the use of the properties of  Bell polynomials to
obtain parameter families of bilinear B\"{a}cklund transformation
for soliton equations.
 As a consequence bilinear B\"{a}cklund transformation with single field can be
 linearize into corresponding Lax pairs  \cite{Gilson}-\cite{Lambert2}.

Our  paper is a further contribution to the theory of Bell
polynomials and supersymmetric equations.  We reconsider Bell
polynomials in a more extended context-- superspace.  We define a
kind of new generalized super Bell polynomials and discuss their
relations with super bilinear equations, which actually provides an
approach to systematically investigate complete integrability of
supersymmetric systems.    As illustrative examples, the bilinear
representations, bilinear B\"{a}cklund transformations, Lax pais and
infinite conservation laws of the supersymmetric KdV equation and
supersymmetric sine-Gordon equation are obtained in a quick and
natural manner.

  The layout  of this paper is as follows.   In Section 2, we briefly recall  elementary
notations about superdifferential, integrals and super  bilinear
operators on superspace. As In Section 3, we propose theory of super
Bell polynomials and establish their connections with supersymmetric
equations.  As consequence a approach to investigate integrability
of supersymmetric equations is presented. In the Sections 4 and 5,
as applications of super Bell polynomials, we
  study integrability of supersymmetric KdV equation supersymmetric sine-Gordon  equation,
 respectively. At last, we  briefly discuss further possible generalization and  applications of Bell polynomials
  and   future work  in Section 6. \\[12pt]
%%%%%%%%%%%%%%%%%%%%%%%%%%%%%%%%%%%%%%%%%%%%%%%%%%%%%%%%%%%%%%%%%%%%%%%%
%%%%%%%%%%%%%%%%%%%%%%%%%%%%%%%%%%%%%%%%%%%%%%%%%%%%%%%%%%%%%%%%%%%%%%%%%%%%%%
{\bf\large  2.   Derivatives and bilinear  operators on superspace}\\

 To make our presentation easily  understanding
and self-contained,  in this section we first briefly review some
notations about superanalysis \cite{Vlad1}-\cite{Vlad2}  and
super-Hirota bilinear operators \cite{Yung, Carstea1}.

A superalgebra  is a $Z_2$-graded space
$\Lambda=\Lambda_0\oplus\Lambda_1$ in which, $\Lambda_0$ is a
subspace consisting of even elements  and $\Lambda_1$ is a subspace
consisting of odd elements. A parity function is introduced for
homogeneous elements on the $\Lambda$, namely, $|a|=0$ if
$a\in\Lambda_0$ and $|a|=1$ if $a\in\Lambda_1$.

The superalgebra  is said to be commutative if the supercommutator
$[a,b]=ab-(-1)^{|a||b|}ba=0$, for arbitrary homogeneous elements $a,
b\in \Lambda$.

 A commutative superalgebra $\Lambda$ with unit $e=1$ is
called a finite-dimensional  Grassmann algebra if it contains a
system of anticommuting generators $\theta_j, j=1, \cdots, n$ with
the anticommutative  property: $[\theta_j,
\theta_k]=\theta_j\theta_k+\theta_k\theta_j=0, \  \theta_j^2=0, \ j,
k=1,2,\cdots, n$.

 Let $\Lambda=\Lambda_0\oplus\Lambda_1$ be a
finite-dimensional Grassmann algebra, then the Banach space
$\mathbb{R}_{\Lambda}^{m,n}=\Lambda_0^m\times\Lambda_1^n$ is called
a superspace of dimension $(m,n)$ over $\Lambda$.  In particular, if
$\Lambda_0=\mathbb{C}$ and $\Lambda_1=0$, then
$\mathbb{R}_{\Lambda}^{m,n}=\mathbb{C}^m.$  We may take even-valued
complex space $\Lambda_0=\mathbb{C}$ in our context.

 A function $f(\boldsymbol{x},\boldsymbol{\theta} ):
\mathbb{R}_{\Lambda}^{m,n}\rightarrow \Lambda$ is said to be
superdifferentiable at the point
$(\boldsymbol{x},\boldsymbol{\theta} )\in
\mathbb{R}_{\Lambda}^{m,n}$ with even coordinates
$\boldsymbol{x}=(x_1, \cdots, x_m)$ and odd coordinates
$\boldsymbol{\theta}=(\theta_1, \cdots, \theta_n)$, if there exist
elements $F_j(\boldsymbol{x},\boldsymbol{\theta} ),
\widetilde{F}_k(\boldsymbol{x},\boldsymbol{\theta} )\in\Lambda, \
j=1, \cdots, m; k=1, \cdots, n$, such that
$$f(\boldsymbol{x}+\boldsymbol{h}, \boldsymbol{\theta}+\boldsymbol{\widetilde{h}})
=f(\boldsymbol{x},\boldsymbol{\theta})+\sum_{j=1}^{m}\langle
F_j(\boldsymbol{x},
\boldsymbol{\theta}),h_j\rangle+\sum_{k=1}^{n}\langle
\widetilde{F}_k(\boldsymbol{x},
\boldsymbol{\theta}),\widetilde{h}_k\rangle+o(||(\boldsymbol{h},
\boldsymbol{\widetilde{h}})||),$$ where the vectors
$\boldsymbol{h}=(h_1, \cdots, h_m)\in\Lambda_0^m$ and
$\boldsymbol{\widetilde{h}}=(\widetilde{h}_{1}, \cdots,
\widetilde{h}_{n})\in\Lambda_1^n$.  The $F_j(\boldsymbol{x},
\boldsymbol{\theta}), \widetilde{F}_k(\boldsymbol{x},
\boldsymbol{\theta})$ are called the super partial derivative of $f$
with respect to $x_j, \theta_k$ at the point $(\boldsymbol{x},
\boldsymbol{\theta})$ and are denoted, respectively, by
$$\frac{\partial f(\boldsymbol{x},
\boldsymbol{\theta})}{\partial x_j}=F_j(\boldsymbol{x},
\boldsymbol{\theta}),\ \ \frac{\partial f(\boldsymbol{x},
\boldsymbol{\theta})}{\partial\theta_k}=F_k(\boldsymbol{x},
\boldsymbol{\theta}), \  j=1,\cdots, m; k=1,\cdots,n.$$ The
derivatives $\frac{\partial f(\boldsymbol{x},
\boldsymbol{\theta})}{\partial x_j}$ with respect to even variables
$x_j, \ j=1,2,\cdots m$ are uniquely defined. While the derivatives
$\frac{\partial f(\boldsymbol{x}, \boldsymbol{\theta})}{\partial
\theta_k}$ to odd variables $\theta_k, \ k=1,2,\cdots n$ are not
uniquely defined, but with an accuracy to within an addition
constant $c\theta_1\cdots\theta_n, c\in \Lambda_0$  from the
annihilator $^\perp L_n$, $L_n=\{\theta_1\cdots\theta_n,
\boldsymbol{\theta}\in \Lambda_1^n\}$.

Let $f=f(\boldsymbol{x},\boldsymbol{\theta}),
g=g(\boldsymbol{x},\boldsymbol{\theta}):
\mathbb{R}_{\Lambda}^{m,n}\rightarrow \Lambda$ be a
superdifferentiable function,   then  super derivative also
satisfies Leibnitz formula
$$\begin{aligned}
&\partial_{x_j} (fg)=(\partial_{x_j} f)g+f(\partial_{x_j}g), \  \ j=1, \cdots, m,\\
&\partial_{\theta_k} (fg)=(\partial_{\theta_k}
f)g+(-1)^{|f|}f(\partial_{\theta_k}g), \ \ k=1, \cdots, n.
\end{aligned}$$

Let differential operators
$\mathcal{D}_k=\partial_{\theta_k}+\theta_k\partial_{x_r}$ ( $k=1,
\cdots, n; r\in\{1,\cdots, m\}$ ) be supersymmetric covariant
derivatives, we can show that they  satisfy
$$\begin{aligned}
& \mathcal{D}_k (fg)=(\mathcal{D}_k
f)g+(-1)^{|f|}f(\mathcal{D}_kg), \\
&[\mathcal{D}_j, \mathcal{D}_k]=0,  \ \
\mathcal{D}_k^2=\partial_{x_r}.
\end{aligned}\eqno(2.1)$$

 Denote by  $\mathcal{P}(\Lambda_1^n,
\Lambda)$ the set of polynomials defined on $\Lambda_1^n$ with value
in $\Lambda$. We say that  a super integral is a map $I:
\mathcal{P}(\Lambda_1^n, \Lambda)\rightarrow \Lambda$ satisfying the
following condition is an super Berezin integral about Grassmann
variables

(1) A linearity: $I(\mu f+\nu g)=\mu I(f)+\nu I(g), \ \mu,
\nu\in\Lambda, \ f, g\in\mathcal{P}(\Lambda_1^n, \Lambda); $

(2) translation invariance: $I(f_{\xi})=I(f)$, where
$f_{\xi}=f(\boldsymbol{\theta}+\boldsymbol{\xi})$ for all
$\boldsymbol{\xi}\in\Lambda_1^n$, $f\in\mathcal{P}(\Lambda_1^n,
\Lambda).$

We denote $I(\theta^\varepsilon)=I_\varepsilon$, where $\varepsilon$
belongs to the set of multiindices
$N_n=\{\boldsymbol{\epsilon}=(\varepsilon_1, \cdots, \varepsilon_n),
\varepsilon_j=0,1,
\boldsymbol{\theta}^\varepsilon=\theta_1^{\varepsilon_1}\cdots\theta_n^{\varepsilon_n}\not\equiv
0\}$. In the case when $I_\varepsilon=0, \varepsilon\in N_n,
|\varepsilon|\leq n=n-1$, such kind of integral has the form
$$I(f)=J(f)I(1,\cdots,1)\equiv \frac{\partial^nf(0)}{\partial\theta_1\cdots\partial\theta_n}I(1,\cdots,1),$$
Since the derivative is defined with an accuracy to with an additive
constant form the annihilator $^\perp L_n$, it follows that
$J:\mathcal{P}\rightarrow \Lambda/^\perp L_n$ is single-valued
mapping. This mapping also satisfies the conditions 1 and 2, and
therefore we shall call it an integral and denote
$$J(f)=\int f(\boldsymbol{\theta})d\boldsymbol{\theta}=\int\theta_1\cdots\theta_n d\theta_1\cdots d\theta_n,$$
which has properties:
$$\begin{aligned}
&\int\theta_1\cdots\theta_n d\theta_1\cdots d\theta_n=1,\\
&\int\frac{\partial f}{\partial\theta_j}d\theta_1\cdots
d\theta_n=0,\ j=1, \cdots, n.\\
&\int f(\boldsymbol{\theta})\frac{\partial
g(\boldsymbol{\theta})}{\partial\theta_j}d\boldsymbol{\theta}=
(-1)^{1+|g|}\int \frac{\partial
f(\boldsymbol{\theta})}{\partial\theta_j}g(\boldsymbol{\theta})d\boldsymbol{\theta}.
\end{aligned}\eqno(2.2)$$

 For  a pair of Grassmann-valued functions
 $f(\boldsymbol{x},\boldsymbol{\theta}), \
g(\boldsymbol{x},\boldsymbol{\theta}):
\mathbb{R}_{\Lambda}^{m,n}\rightarrow \Lambda$, the  ordinary Hirota
bilinear operator is defined by
$$\begin{aligned}
& D_{x_j}f(\boldsymbol{x},\boldsymbol{\theta})\cdot
g(\boldsymbol{x},\boldsymbol{\theta})=(\partial_{x_j}-\partial_{x_j'})
f(\boldsymbol{x},\boldsymbol{\theta})g(\boldsymbol{x}',\boldsymbol{\theta}')|_{\boldsymbol{x'=x,
\theta'=\theta}}, \ j=1,\cdots, m,
\end{aligned}$$
and  super-Hirota bilinear operators are defined as
$$\begin{aligned}
& S_kD_{x_j}f(\boldsymbol{x},\boldsymbol{\theta})\cdot
g(\boldsymbol{x},\boldsymbol{\theta})=(\mathcal{D}_k-\mathcal{D}_{k}')
f(\boldsymbol{x},\boldsymbol{\theta})g(\boldsymbol{x}',\boldsymbol{\theta}')|_{\boldsymbol{x'=x,
\theta'=\theta}}, \ \  k=1,\cdots,n,
\end{aligned}$$
here we have denoted
$\mathcal{D}_{k}'=\partial_{\theta'}+\theta'\partial_x$.

It can be shown that these super-Hirota bilinear
  operators have  properties
$$\begin{aligned}
&S_k^{2N}f\cdot g=D_{x_r}^Nf\cdot g,\ \ N\in \mathbb{Z}\\
&S_kf\cdot g=(\mathcal{D}_k f)g-(-1)^{|f|}f(\mathcal{D}_k g).
\end{aligned}$$

In our context, we are interested in  bosonic, also called even,
superfield  function $f(\boldsymbol{x},\boldsymbol{\theta}):
\mathbb{R}_{\Lambda}^{m,n} \rightarrow
\mathbb{R}_{\Lambda}^{1,0}=\Lambda_0 $.  It can be expanded in
powers of odd coordinates $\theta_k, \ k=1, \cdots, n$, that is,
$$f=f_0(\boldsymbol{x})+\sum_{k\geq 0}\sum_{j_1<\cdots<j_k}f_{j_1\cdots j_k}(\boldsymbol{x})\theta_{j_1}\cdots\theta_{j_k},$$
where the coefficients $f_{j_1\cdots j_k}(\boldsymbol{x})\in
\Lambda_0$ are even
functions with respect to $x_1, \cdots, x_m$.\\[12pt]
%%%%%%%%%%%%%%%%%%%%%%%%%%%%%%%%%%%%%%%%%%%%%%%%%%%%%%%%%%%%%%%%%%%%%%%%%%%%
{\bf\large 3. Generalized super Bell polynomials on superspace}

 Based on the above fundamental notations, in this section we develop theory  of generalized super Bell
polynomials, which are a main tool to study  the integrability of
supersymmetric equations.
\\[12pt]
%%%%%%%%%%%%%%%%%%%%%%%%%%%%%%%%%%%%%%%%%%%%%%%%%%%%%%%%%%%%%%%%%%%%
{\bf 2.1.  Generalized  super Bell polynomials}

 To well compare our super Bell polynomials with ordinary ones,
 let's first simply  recall the rdinary Bell polynomials.  During the early 1930s, Bell
introduced three kinds of exponential polynomials \cite{Bell}.

 The first Bell polynomials are defined as
$$ \xi_{n}(x, t, r)= e^{-tx^{r}}\partial_{x}^{n} e^{tx^{r}},\eqno(3.1)$$
where  $r>0$ is a constant integer, $n\geq 0$ an arbitrary integer,
and $x, t\in \mathbb{R}$ independent variables. For $r=2$, the Bell
polynomials $\xi_n(x,t)$ are exactly Hermite polynomials.

The Bell polynomials are algebraic polynomials in two elements $x$
and $t$.  The first few lowest order Bell Polynomials are
$$\begin{aligned}
&\xi_0 (x,t,r)=1, \ \ \xi_1(x,t,r)=rtx^{r-1},\ \ \xi_2(x,t,r)=r^2t^2x^{2r-2}+r(r-1)tx^{r-2},\\
&
\xi_3(x,t,r)=r^3t^3x^{3r-3}+3r^2(r-1)t^2x^{2r-3}+r(r-1)(r-2)tx^{r-3}.
\end{aligned}$$

The second  Bell polynomials are a  generalization of  the
 Bell polynomials (3.1) and defined by
$$
\phi_n=\phi(\alpha_1,\cdots,\alpha_n),\ \ \phi_0=1,\ \
\phi_{n+1}=\sum_{s=1}^n\left(\begin{matrix}n\cr
s\end{matrix}\right)\alpha_{s+1}\phi_{n-s}, $$ where
$(\alpha_1,\cdots,\alpha_n,\cdots)$ is an infinite sequence of
independent variables.  For the particular sequence
$\alpha_j=j!\left(\begin{matrix}\ell_i\cr
r_i\end{matrix}\right)xt^{r-j}, \ j=1, \cdots, r; \ \alpha_j=0, \
j>r$, we have
$$\phi_n=\xi_n(x,t,r).$$

The Bell polynomial $\phi_n$  is a polynomial about variables
$\alpha_1,\cdots,\alpha_n$. For instance,  the first three  of
second Bell polynomials read
$$\begin{aligned}
&\phi_0=1, \ \ \phi_1=\alpha_1^2+\alpha_2,\ \
\phi_3=\alpha_1^3+3\alpha_1\alpha_2+\alpha_3.
\end{aligned}$$

  The third   Bell polynomials, further generalization of the $\xi_n$ and $\phi_n$,  are defined by
$$\begin{aligned}
& Y_n=Y_n(y_t,\cdots, y_{nt})=e^{-y}\partial_t^ne^{y},
\end{aligned}\eqno(3.2)$$
where  $y=e^{\alpha
t}-\alpha_0\equiv\alpha_1t+\alpha_2t^2/2!+\cdots$, and  we have
denoted derivative notation  $y_{kt}=\partial_t^k y$.
 For the spacial case when  $\alpha_r=r!x, \
\alpha_k=0, \ k\not=r$, then we have
$$Y_n=\xi_n(x,t,r).$$

The polynomials (3.3)   are  polynomials about the derivatives  of
function $y$, for example,  the first three are
$$\begin{aligned}
&Y_0=1, \ \ Y_1=y_{t}, \ Y_2=y_{t}^2+y_{2t},\ \
Y_3=y_{t}^3+3y_ty_{2t}+y_{3t},
\end{aligned}$$

 More recently  Lambert et al  generalized
 the third Bell polynomials as
$$\begin{aligned}
& Y_n=e^{-y}\partial_{x_1}^{n_1}\cdots\partial_{x_m}^{n_m} e^{y},
\end{aligned}\eqno(3.3)$$
where  $y=y(x_1,\cdots,x_m): \mathbb{R}^m\rightarrow \mathbb{R}$
\cite{Gilson}-\cite{Lambert2}.

 We now propose  the following  multi-dimensional and super extension
 to   the ordinary Bell polynomials (3.1)-(3.3).

{\bf Definition 1.}  Let $f=f(\boldsymbol{x},\boldsymbol{\theta}):
\mathbb{R}_{\Lambda}^{m,n} \rightarrow \Lambda_0 $ be a
superdifferential bosonic function, the generalized super Bell polynomials
(super $Y$-polynomials) is defined as follows
$$Y_{\boldsymbol{\ell}\cdot\boldsymbol{x},\boldsymbol{\theta}}(f)=Y_{\boldsymbol{\ell}\cdot\boldsymbol{x},\boldsymbol{\theta}}[f_{\boldsymbol{r}\cdot\boldsymbol{x},
\boldsymbol{\mu}\cdot\boldsymbol{\theta}}]\equiv
e^{-f}\mathcal{D}_1\cdots
\mathcal{D}_n\partial_{x_1}^{\ell_1}\cdots\partial_{x_m}^{\ell_m}
e^{f},\eqno(3.4)$$
 where   $\ell_j\geq
0, \ j=1, \cdots, m$ denote arbitrary integers. To make subscript in
expressions simple, we  use  some abbreviation notations in our
context, for example,
$$\begin{aligned}
&\boldsymbol{\ell}=(\ell_1,\cdots, \ell_m), \ \
\boldsymbol{\theta}=(\theta_1,\cdots,\theta_n),\ \
\boldsymbol{\ell}\cdot\boldsymbol{x}=(\ell_1x_1,\cdots,\ell_m x_m),\\
&\boldsymbol{r}=(r_1,\cdots, r_m), \ \
\boldsymbol{r}\cdot\boldsymbol{x}=(r_1x_1,\cdots,r_m x_m),\\
 &\boldsymbol{\mu}=(\mu_1,\cdots, \mu_n), \ \
\boldsymbol{\mu}\cdot\boldsymbol{\theta}=(\mu_1\theta_1,\cdots,\mu_n
\theta_n).
\end{aligned}$$

{\bf Remark 1.} The first notation
$Y_{\boldsymbol{\ell}\cdot\boldsymbol{x},\boldsymbol{\theta}}(f) $
in (3.4)  denotes the   $\ell_j$-order derivatives  of $f$ with
respect to the variable $x_j, j=1,\cdots, m$ and  covariant
derivatives with respect to ${\theta_k}, \  k=1,\cdots, n$. The
second notation
$Y_{\boldsymbol{\ell}\cdot\boldsymbol{x},\boldsymbol{\theta}}[f_{\boldsymbol{r}\cdot\boldsymbol{x},
\boldsymbol{\mu}\cdot\boldsymbol{\theta}}]$ implies that the super
Bell polynomials (3.4) should be understood as such a multivariable
differential polynomial with respect to  partial derivatives
$f_{\boldsymbol{r}\cdot\boldsymbol{x},
\boldsymbol{\mu}\cdot\boldsymbol{\theta}}$ ( $ r_j=0,\cdots,\ell_j,
\ j=1,\cdots, m, \mu_k=0,1, k=1, \cdots, n$),  but not variable
elements $x_j, \theta_k \ ( j=1, \cdots, m; k=1, \cdots, n)$ as
ordinary Bell polynomials. For instance, the ${Y}_{3x}(f)$ in the
next  example is a polynomial $Y_{3x}(f_x, f_{2x}, f_{3x})$ with
respect to three variable elements $f_x, f_{2x}, f_{3x}$.

 To better understanding our generalized super Bell
polynomials, let us see an illustrative example. For the special
case $f=f(x, \theta_1, \theta_2)$, the associated super Bell
polynomials defined by (3.4) read
$$\begin{aligned}
&{Y}_x(f) =f_x, \ \ \ {Y}_{2x}(f)=f_{2x}+f_x^2,\\
&{Y}_{3x}(f)=f_{3x}+3f_xf_{2x}+f_x^3,\ \
Y_{\theta_1}(f)=\mathcal{D}_1f, \\
&Y_{\theta_1\theta_2}(f)=\mathcal{D}_1\mathcal{D}_2f+(\mathcal{D}_1f)\mathcal{D}_2f,
\ \ {Y}_{x,\theta_1}(f) =\mathcal{D}_1f_{x}+f_x\mathcal{D}_1f, \\
&{Y}_{2x,\theta_1}(f)=f_{2x}\mathcal{D}_1f+\mathcal{D}_1f_{2x}+f_x^2\mathcal{D}_1f+2f_{x}\mathcal{D}_1f_x,\\
&{Y}_{3x,\theta_1}(f)=f_{3x}\mathcal{D}_1f+3f_xf_{2x}\mathcal{D}_1f+3f_{2x}\mathcal{D}_1f_x+3f_{x}\mathcal{D}_1f_{2x}+3f_x^2\mathcal{D}_1f_x+\mathcal{D}_1f_{3x}.
\end{aligned}$$

Let see the relations between our generalized  super polynomials and
ordinary Bell polynomials, as well as ordinary generalized Bell
polynomials.

For the special case $\mathbb{R}_{\Lambda}^{m,n}=\mathbb{R}^2,
 \ell_2=0$, $f=f(x_1, x_2)=x_2x_1^r$ with the
constant integer $r>0$,  then (3.4) reduces to the first Bell
polynomials (3.1)
$$Y_{\ell_1 x_1}(f)=e^{{-x_2x_1^{r}}}\partial_{x_1}^{\ell_1}e^{{x_2x_1^{r}}}=\xi_{\ell_1}(x_1, x_2).$$

For the case $\mathbb{R}_{\Lambda}^{m,n}=\mathbb{R}^m$, the
corresponding  generalized super Bell polynomials (3.4) degenerates
to  generalized Bell polynomials (3.3) given by Lambert et al. The
Bell polynomials  admit  partitional representation \cite{Gilson}
$$Y_{\boldsymbol{\ell}\cdot\boldsymbol{x}}(f)=\sum\frac{\ell_1!\cdots\ell_m!}{c_1!\cdots
c_k!}\prod_{j=1}^{k}\left(\frac{f_{r_{1j},\cdots,r_{mj}}}{r_{1j}!\cdots
r_{mj}!}\right)^{c_j},\eqno(3.5)$$ where the sum is to taken over
all partitions $[(r_{j1},\cdots, r_{m1})^{c_1}, \cdots,
(r_{1k},\cdots, r_{mk})^{c_k}]$ the $m$-tuple $(\ell_1, \cdots,
\ell_m)$.

In following, we investigate  properties of super Bell polynomials
which are  key results to establish connects with  supersymmetric
equations.

{\bf Theorem 1.}  Under the Hopf-Cole transformation $f=\ln \psi$,
the generalized  super Bell polynomials $Y_{\boldsymbol{r}\cdot
\boldsymbol{x},\boldsymbol{\mu}\cdot\boldsymbol{\theta}}(f)$ can be
``linearized" into the form
$$Y_{\boldsymbol{r}\cdot
\boldsymbol{x},\boldsymbol{\mu}\cdot\boldsymbol{\theta}}(f)|_{f=\ln
\psi}= \psi_{\boldsymbol{r}\cdot
\boldsymbol{x},\boldsymbol{\mu}\cdot\boldsymbol{\theta}}/\psi.\eqno(3.6)$$

{\it Proof.}  According to the definition (3.4), we have
$$\begin{aligned}
&Y_{\boldsymbol{r}\cdot
\boldsymbol{x},\boldsymbol{\mu}\cdot\boldsymbol{\theta}}(f)|_{f=\ln\psi}=e^{-\ln
\psi}\mathcal{D}_1^{\mu_1}\cdots
\mathcal{D}_n^{\mu_n}\partial_{x_1}^{r_1}\cdots\partial_{x_m}^{r_m}
e^{\ln\psi} =\psi_{\boldsymbol{r}\cdot
\boldsymbol{x},\boldsymbol{\mu}\cdot\boldsymbol{\theta}}/{\psi},
\end{aligned}$$
which finishes the proof of Theorem 1. $\square$

{\bf Remark 2.} According to the theorem, under the Hopf-Cole
transformation $f=\ln \psi$, a  equation in term of linear
combination of super Bell polynomials, i.e.
$$\sum_{\boldsymbol{r},\boldsymbol{\mu}}C_{\boldsymbol{r},\boldsymbol{\mu}}(\boldsymbol{x}, \boldsymbol{\theta}) Y_{\boldsymbol{r}\cdot
\boldsymbol{x},\boldsymbol{\mu}\cdot\boldsymbol{\theta}}(f)=0$$ can
be linearized into the form
$$\sum_{\boldsymbol{r},\boldsymbol{\mu}}C_{\boldsymbol{r},\boldsymbol{\mu}}(\boldsymbol{x}, \boldsymbol{\theta}) \psi_{\boldsymbol{r}\cdot
\boldsymbol{x},\boldsymbol{\mu}\cdot\boldsymbol{\theta}}=0,$$ where
$C_{\boldsymbol{r},\boldsymbol{\mu}}(\boldsymbol{x},
\boldsymbol{\theta})$ are functions independent of the function $f$.
This is a key property to  construct the Lax pair of supersymmetric
equations.

{\bf Theorem 2. }  The  super  Bell polynomials (3.4) admit
recursion formula
$$Y_{\boldsymbol{\ell}\cdot \boldsymbol{x},\boldsymbol{\theta}}(f)=\prod_{k=1}^{n}(\mathcal{D}_k
+\mathcal{D}_kf)Y_{\boldsymbol{\ell}\cdot
\boldsymbol{x}}(f).\eqno(3.7)$$

{\it Proof.} By  the definition (3.4),  direct computation leads to
$$\begin{aligned}
&Y_{\boldsymbol{\ell}\cdot
\boldsymbol{x},\boldsymbol{\theta}}(f)=\mathcal{D}_1Y_{\boldsymbol{\ell}\cdot
\boldsymbol{x},\theta_2,\cdots,\theta_n}(f)+
(\mathcal{D}_1f)Y_{\boldsymbol{\ell}\cdot
\boldsymbol{x},\theta_2,\cdots,\theta_n}(f)\\
&=(\mathcal{D}_1+\mathcal{D}_1f)Y_{\boldsymbol{\ell}\cdot
\boldsymbol{x},\theta_2,\cdots,\theta_n}(f).
\end{aligned}$$
Similarly,
$$\begin{aligned}
&Y_{\boldsymbol{\ell}\cdot
\boldsymbol{x},\theta_2,\cdots,\theta_n}(f)=(\mathcal{D}_2+\mathcal{D}_2f)Y_{\boldsymbol{\ell}\cdot
\boldsymbol{x},\theta_3,\cdots,\theta_n}(f).
\end{aligned}$$
Repeating the above arguments  then proves the formula (3.7).
$\square$

{\bf Theorem  3.}  The super  Bell polynomials (3.4) possess parity
property
$$\begin{aligned}
&Y_{\boldsymbol{\ell}\cdot \boldsymbol{x},\boldsymbol{\theta}}[
(-1)^{ \sum r_j+ \sum \mu_k}f_{\boldsymbol{r}\cdot \boldsymbol{x},
\boldsymbol{\mu}\cdot\boldsymbol{\theta}}]=(-1)^{n+\sum\ell_j}Y_{\boldsymbol{\ell}\cdot
\boldsymbol{x}, \boldsymbol{\theta}}[f_{\boldsymbol{r}\cdot
\boldsymbol{x},\boldsymbol{\mu}\cdot\boldsymbol{\theta}}].\end{aligned}\eqno(3.8)$$

{\it Proof.}  From  the recursion relation (3.7),  it follows that
$$\begin{aligned}
& Y_{\boldsymbol{\ell}\cdot
\boldsymbol{x},\boldsymbol{\theta}}[(-1)^{\footnotesize \sum r_j+
\sum \mu_k}f_{\boldsymbol{r}\cdot \boldsymbol{x},
\boldsymbol{\mu}\cdot\boldsymbol{\theta}}]\stackrel{}{=}\prod_{k=1}^{n}(-\mathcal{D}_k-\mathcal{D}_kf)Y_{\boldsymbol{\ell}\cdot
\boldsymbol{x}}[(-1)^{ \sum r_j}f_{\boldsymbol{r}\cdot \boldsymbol{x}}]\\
&=(-1)^n\prod_{k=1}^{n}(\mathcal{D}_k+\mathcal{D}_kf)Y_{\boldsymbol{\ell}\cdot
\boldsymbol{x}}[(-1)^{ \sum r_j}f_{\boldsymbol{r}\cdot
\boldsymbol{x}}].
\end{aligned}\eqno(3.9)$$
 While applying the  partitional representation (3.5),  we have
$$\begin{aligned}
& Y_{\boldsymbol{\ell}\cdot \boldsymbol{x}}[(-1)^{ \sum
r_j}f_{\boldsymbol{r}\cdot \boldsymbol{x}}]\stackrel{}{=}(-1)^{\sum
\ell_j}Y_{\boldsymbol{\ell}\cdot
\boldsymbol{x}}[f_{\boldsymbol{r}\cdot \boldsymbol{x}}].
\end{aligned}\eqno(3.10)$$
Hence, combing (3.7), (3.9)  and  (3.10) proves  the formula (3.8).
$\square$

{\bf Theorem 4.}   The super  Bell polynomials (3.4) obey addition
property
$$\begin{aligned}
&Y_{\boldsymbol{\ell}\cdot
\boldsymbol{x},\boldsymbol{\theta}}(f+g)=\sum_{\mu_1,\cdots,
\mu_n=0}^1 (-1)^{\tau[\{\boldsymbol{(1-\mu)\cdot
n},\boldsymbol{\mu\cdot n}\}\setminus \{0\}]}
\sum_{r_1=0}^{\ell_1}\cdots\sum_{r_m=0}^{\ell_m}\prod_{i=1}^{m}
\left(\begin{matrix}\ell_i\cr
r_i\end{matrix}\right)\\
&\times Y_{\boldsymbol{(\ell-r)\cdot
x},\boldsymbol{(1-\mu)\cdot\theta}}(f)Y_{\boldsymbol{r\cdot
x},\boldsymbol{\mu\cdot\theta}}(g),\end{aligned}\eqno(3.11)$$
 where $\boldsymbol{n}=(1, 2, \cdots, n)$;  $\tau[\{\boldsymbol{(1-\mu)\cdot n},\boldsymbol{\mu\cdot
n}\}\setminus \{0\}]$ denotes the reverse order numbers of the
$n$-order permutation $\{\boldsymbol{(1-\mu)\cdot
n},\boldsymbol{\mu\cdot n}\}\setminus \{0\}$, which is generated
from anticommutation of covariant derivatives $\mathcal{D}_j, \ j=1,
\cdots, n$, and obtained from a $2n$-order permutation
$\{(1-\mu_1)1, \cdots, (1-\mu_n)n, \mu_1 1, \cdots, \mu_n n\}$
($\mu_j=0$ or $1$) by taking off all zero terms. The $2n$-order
permutation is the subscript of corresponding covariant derivatives
of  term $Y_{\boldsymbol{(\ell-r)\cdot
x},\boldsymbol{(1-\mu)\cdot\theta}}(f)Y_{\boldsymbol{r\cdot
x},\boldsymbol{\mu\cdot\theta}}(g)$ kept in original order.

{\it Proof.}   According  to the commutative properties of covariant
derivatives (3.1),   a minus sign in the  Leibnitz rule  exactly
corresponds to an inverse order of the $n$-order permutation $\{1,
2, \cdots, n\}$.  So direct computation shows that
$$\begin{aligned}
&(FG)^{-1}\mathcal{D}_1\cdots\mathcal{D}_n\partial_{x_1}^{\ell_1}\cdots
\partial_{x_m}^{\ell_m}(FG)=\sum_{\mu_1,\cdots,
\mu_n=0}^1  (-1)^{\tau[\{\boldsymbol{(1-\mu)\cdot
n},\boldsymbol{\mu\cdot n}\}\setminus \{0\}]}
\sum_{r_1=0}^{\ell_1}\cdots\sum_{r_m=0}^{\ell_m}\prod_{i=1}^{m}
\left(\begin{matrix}\ell_i\cr
r_i\end{matrix}\right)\\
&\times
\left(F\mathcal{D}_1^{1-\mu_1}\cdots\mathcal{D}_n^{1-\mu_n}\partial_{x_1}^{\ell_1-r_1}\cdots
\partial_{x_m}^{\ell_m-r_m}F\right)\left(G\mathcal{D}_1^{\mu_1}\cdots\mathcal{D}_n^{\mu_n}\partial_{x_1}^{r_1}\cdots
\partial_{x_m}^{r_m}G\right),
\end{aligned}$$
which implies (3.11) by replacing $F=e^{f}$ and $ G=e^{g}$.
$\square$

Let $F, G: \mathbb{R}_{\Lambda}^{m,n} \rightarrow \Lambda_0 $ be two
bosonic superdifferential  functions, then direct computation yields
$$\begin{aligned}
&\mathcal{D}_1\mathcal{D}_2\mathcal{D}_3(FG)=(\mathcal{D}_1\mathcal{D}_2\mathcal{D}_3F)G+
(\mathcal{D}_2\mathcal{D}_3F)\mathcal{D}_1G-(\mathcal{D}_1\mathcal{D}_3F)\mathcal{D}_2G\\
&+(\mathcal{D}_2F)(\mathcal{D}_1F)\mathcal{D}_2G+(\mathcal{D}_1\mathcal{D}_2F)\mathcal{D}_3G
-(\mathcal{D}_2F)(\mathcal{D}_1F)\mathcal{D}_3G\\
&+(\mathcal{D}_1F)(\mathcal{D}_2F)\mathcal{D}_3G+F(\mathcal{D}_1\mathcal{D}_2\mathcal{D}_3G),
\end{aligned}$$
in which corresponding six  even permutations are $\{1,2,3\},
\{2,3,1\}, \{3,1,2\}, \{1,2,3\}$, $\{1,2,3\}, \{1,2,3\}$, and two
odd permutations are $\{1,3,2\}, \{2,1,3\} $.
\\[12pt]
%%%%%%%%%%%%%%%%%%%%%%%%%%%%%%%%%%%%%%%%%%%%%%%%%%%%%%%%%%%%%%%%%%%%
{\bf 2.2. Generalized super binary Bell polynomials}

We further define a class  of  super binary  Bell polynomials which
play an important role in the study of integrability for
supersymmetric equations.

 {\bf Definition 2. }  Based  on the use of above super Bell polynomials (3.4), the  super binary
Bell polynomials ( $\mathcal{Y}$-polynomials)   can be defined as
follows
$$\mathcal{Y}_{\boldsymbol{\ell}\cdot \boldsymbol{x},\boldsymbol{\theta}}(v,w)=Y_{\boldsymbol{\ell}\cdot \boldsymbol{x},\boldsymbol{\theta}}
[f_{\boldsymbol{r}\cdot
\boldsymbol{x},\boldsymbol{\mu}\cdot\boldsymbol{\theta}}],\eqno(3.12)$$
in which we replace the function $f$ and its derivatives  by
corresponding terms of  functions $w$ and $v$ respectively,
according the following rule
$${f_{\boldsymbol{r}\cdot \boldsymbol{x},\boldsymbol{\mu}\cdot\boldsymbol{\theta}}
=\left\{\begin{matrix}v_{\boldsymbol{r}\cdot
\boldsymbol{x},\boldsymbol{\mu}\cdot\boldsymbol{\theta}},& {\rm if}\
\sum_{j=1}^{m} r_j+\sum_{k=1}^{n}\mu_k\  {\rm is\ \ odd},\cr\cr
w_{\boldsymbol{r}\cdot
\boldsymbol{x},\boldsymbol{\mu}\cdot\boldsymbol{\theta}},&{\rm if}\
\sum_{j=1}^{m} r_j+\sum_{k=1}^{n}\mu_k\ \ {\rm is\ \
even},\end{matrix}\right.}$$ The  super binary Bell polynomials
(3.6) is   multi-variable polynomials with respect to various
partial derivatives $v_{\boldsymbol{r}\cdot
\boldsymbol{x},\boldsymbol{\mu}\cdot\boldsymbol{\theta}}$ and
$w_{\boldsymbol{r}\cdot
\boldsymbol{x},\boldsymbol{\mu}\cdot\boldsymbol{\theta}}$,  $
r_j=0,\cdots, \ell_j,\ j=0, \cdots, m, \mu_k=0, 1, k=1, \cdots, n$.

The super binary  Bell polynomials  also inherit the easily
recognizable partial structure of the super Bell polynomials. The
first few  are explicitly calculated as
 $$\begin{aligned}
&\mathcal{Y}_x (v)=v_x, \ \ \mathcal{Y}_{2x}(v,w)=w_{2x}+v_x^2,\ \
 \mathcal{Y}_{3x}(v, w)=v_{3x}+3v_xw_{2x}+v_x^3,\\
&\mathcal{Y}_{\theta_1}(v)=\mathcal{D}_1v, \ \
\mathcal{Y}_{\theta_1\theta_2}(w,
v)=\mathcal{D}_1\mathcal{D}_2w+(\mathcal{D}_1v)\mathcal{D}_2v,\\
&\mathcal{Y}_{x,\theta_1}(v,w) =\mathcal{D}_1w_{x}+v_x\mathcal{D}_1v,\\
&\mathcal{Y}_{2x,\theta_1}(v,w)=w_{2x}\mathcal{D}_1v+\mathcal{D}_1v_{2x}+v_x^2\mathcal{D}_1v+2v_x\mathcal{D}_1w_x,\\
&\mathcal{Y}_{3x,\theta_1}(v,w)=v_{3x}\mathcal{D}_1v+3v_xw_{2x}\mathcal{D}_1v+3w_{2x}\mathcal{D}_1w_x+3v_{x}\mathcal{D}_1v_{2x}\\
&+3v_x^2\mathcal{D}_1w_x+\mathcal{D}_1w_{3x}.
\end{aligned}\eqno(3.13)$$

We denote the special case of super Bell polynomials by
$\mathcal{Y}_{\boldsymbol{\ell}\cdot
\boldsymbol{x},\boldsymbol{\theta}}(v=0,w)=P_{\boldsymbol{\ell}\cdot
\boldsymbol{x},\boldsymbol{\theta}}(w)$, then   it follows from
(3.13) that
 $$\begin{aligned}
&P_{2x}(w)=w_{2x}, \  \  P_{4x}(w)=w_{4x}+3w_{2x}^2,\ \
P_{\theta_1,\theta_2}(w)=\mathcal{D}_1\mathcal{D}_2w,\\
&P_{x,\theta_1}(w)=\mathcal{D}_1w_x, \ \
P_{3x,\theta_1}(w)=\mathcal{D}_1w_{3x}+3w_{2x}\mathcal{D}_1w_x,
\cdots.
\end{aligned}\eqno(3.14)$$

{\bf Theorem 5.}  The link between super binary  Bell polynomials
$\mathcal{Y}_{\boldsymbol{\ell}\cdot
\boldsymbol{x},\boldsymbol{\theta}}(v,w)$ and the super Hirota
bilinear equation $S_1\cdots S_nD_{x_1}^{\ell_1}\cdots
D_{x_m}^{\ell_m}F\cdot G$ can be established  by an identity
$$\begin{aligned}
&\mathcal{Y}_{\boldsymbol{\ell}\cdot
\boldsymbol{x},\boldsymbol{\theta}}(v=\ln F/G, w=\ln FG)=
(FG)^{-1}S_1\cdots S_nD_{x_1}^{\ell_1}\cdots D_{x_m}^{\ell_m} F\cdot
G.
\end{aligned}\eqno(3.15)$$
This formula will be sued to obtain bilinear B\"{a}cklund
transformations of supersymmetric equations.

{\it Proof.}  Let $f=\ln F, \ g=\ln G$, then we have  $v=f-g,
w=f+g$.  It follows  from  the definition 2 that
$$\begin{aligned}
&\mathcal{Y}_{\boldsymbol{\ell}\cdot
\boldsymbol{x},\boldsymbol{\theta}}(v=\ln F/G, w=\ln
FG)=Y_{\boldsymbol{\ell}\cdot
\boldsymbol{x},\boldsymbol{\theta}}[f_{\boldsymbol{s}\cdot\boldsymbol{x},\boldsymbol{\tilde{\mu}}\boldsymbol{\theta}}
+(-1)^{\sum s_i+\sum \tilde{\mu}_j} g_{\boldsymbol{s}\cdot\boldsymbol{x},\boldsymbol{\tilde{\mu}}\boldsymbol{\theta}}]\\
&\stackrel{(3.11)}{=}\sum_{\mu_1,\cdots, \mu_n=0}^1
(-1)^{\tau[\{\boldsymbol{(1-\mu)\cdot n},\boldsymbol{\mu\cdot
n}\}/0]}
\sum_{r_1=0}^{\ell_1}\cdots\sum_{r_m=0}^{\ell_m}\prod_{i=1}^{m}
\left(\begin{matrix}\ell_i\cr
r_i\end{matrix}\right)\\[8pt]
&\ \ \ \ \times  Y_{\boldsymbol{(\ell-r)\cdot
x},\boldsymbol{(1-\mu)\cdot\theta}}[f_{\boldsymbol{s}\cdot\boldsymbol{x},
\boldsymbol{\tilde{\mu}}\boldsymbol{\theta}}]Y_{\boldsymbol{r\cdot
x},\boldsymbol{\mu\cdot\theta}}[(-1)^{\sum s_i+\sum \tilde{\mu}_j}
g_{\boldsymbol{s}\cdot\boldsymbol{x},\boldsymbol{\tilde{\mu}}\boldsymbol{\theta}}]\\[8pt]
&\stackrel{(3.8)}{=}\sum_{\mu_1,\cdots, \mu_n=0}^1
(-1)^{\tau[\{\boldsymbol{(1-\mu)\cdot n},\boldsymbol{\mu\cdot
n}\}/0]+\sum_{j=1}^m r_j+\sum_{k=1}^n {\mu}_k}
\sum_{r_1=0}^{\ell_1}\cdots\sum_{r_m=0}^{\ell_m}\prod_{i=1}^{m}
\left(\begin{matrix}\ell_i\cr
r_i\end{matrix}\right)\\[4pt]
&\ \ \ \ \times Y_{\boldsymbol{(\ell-r)\cdot
x},\boldsymbol{(1-\mu)\cdot\theta}}(f)Y_{\boldsymbol{r\cdot
x},\boldsymbol{\mu\cdot\theta}}(g)\\
 &=(FG)^{-1}S_1\cdots
S_nD_{x_1}^{\ell_1}\cdots D_{x_m}^{\ell_m} F\cdot G.
\end{aligned}$$ $\square$

 For the particular case
when $F=G$,   the formula (3.12)  reduces to
 $$\begin{aligned}
&G^{-2}S_1\cdots S_nD_{x_1}^{\ell_1}\cdots D_{x_m}^{\ell_m} G\cdot
G=\mathcal{Y}_{\boldsymbol{\ell}\cdot
\boldsymbol{x},\boldsymbol{\theta}}(0,w=2\ln
G)\\[6pt]
&=\left\{\begin{matrix}0,&n+\sum_{j=1}^{m} \ell_j \ \ {\rm is\ \
odd},\cr\cr P_{\boldsymbol{\ell}\cdot
\boldsymbol{x},\boldsymbol{\theta}}(w),&n+\sum_{j=1}^{m} \ell_j \ \
{\rm is\ \ even},\end{matrix}\right.\end{aligned}\eqno(3.16)$$ which
implies that the $P$-polynomials can be  characterized by an equally
recognizable even part partitional structure.  The formulae (3.15)
and (3.16) will prove particularly useful in connecting
supersymmetric  equations with their corresponding super bilinear
equations. Once a nonlinear equation is expressible as a linear
combination of super Bell $\mathcal{Y}$-polynomials or
$P$-polynomials,  then it can be transformed into a super  linear
equation.

{\bf Theorem 6.}  The super  binary Bell polynomials
$\mathcal{Y}_{\boldsymbol{\ell\cdot x},\boldsymbol{\theta}}(v,w)$
can be separated into super $P$-polynomials and  super
 Bell  $Y$-polynomials
$$\begin{aligned}
&\mathcal{Y}_{\boldsymbol{\ell\cdot x},\boldsymbol{\theta}}(v,
w)=\sum_{\mu_1, \cdots, \mu_n=0}^1 (-1)^{\large
\tau[\{\boldsymbol{(1-\mu)\cdot n},\boldsymbol{\mu\cdot n}\}/0]}
\sum_{r_1=0}^{\ell_1}\cdots\sum_{r_m=0}^{\ell_m} \prod_{i=1}^{m}
\left(\begin{matrix}\ell_i\cr
r_i\end{matrix}\right)\\[8pt]
&\ \ \ \times P_{\boldsymbol{r\cdot x},
\boldsymbol{\mu\cdot\theta}}(w-v)Y_{\boldsymbol{(\ell-r)\cdot x},
\boldsymbol{(1-\mu)\cdot\theta}}(v),
\end{aligned}\eqno(3.17)$$
where only non-vanishing contributions being those for which $\sum
r_j+\sum \mu_k$ is even integer.

{\it Proof.}   According Definition 2 of the super Bell polynomials,
we have
$$\mathcal{Y}_{\boldsymbol{p\cdot
x},\boldsymbol{\nu\cdot\theta}}(v=0, w)=0, \ \ {\rm as } \ \ \sum
p_j+\sum\nu_k \ {\rm is\ odd}, $$ so that by using  Theorem 4,
$$\begin{aligned}
&\mathcal{Y}_{\boldsymbol{\ell\cdot x},\boldsymbol{\theta}}(v,
w)=\mathcal{Y}_{\boldsymbol{\ell\cdot x},\boldsymbol{\theta}}(v,
v+q)={Y}_{\boldsymbol{\ell\cdot x},\boldsymbol{\theta}}(
v+q)|_{\large q_{\boldsymbol{p\cdot
x},\boldsymbol{\nu\cdot\theta}}=0}\\[6pt]
 &=\sum_{\mu_1, \cdots, \mu_n=0}^1
(-1)^{\large \tau[\{\boldsymbol{(1-\mu)\cdot n},\boldsymbol{\mu\cdot
n}\}/0]} \sum_{r_1=0}^{\ell_1}\cdots\sum_{r_m=0}^{\ell_m}
\prod_{i=1}^{m} \left(\begin{matrix}\ell_i\cr
r_i\end{matrix}\right)\\[8pt]
&\times \ \ \ Y_{\boldsymbol{r\cdot x},
\boldsymbol{\mu\cdot\theta}}(q) Y_{\boldsymbol{(\ell-r)\cdot x},
\boldsymbol{(1-\mu)\cdot\theta}}(v)|_{\large q_{\boldsymbol{p\cdot
x},\boldsymbol{\nu\cdot\theta}}=0},
\end{aligned}\eqno(3.18)$$
where $q=w-v$, the sum   $\sum p_j+\sum\nu_k $ is odd integer.

Substituting the  relation
$$\begin{aligned}
&Y_{\boldsymbol{r\cdot x}, \boldsymbol{\mu\cdot\theta}}(q)|_{
q_{\boldsymbol{p\cdot x},\boldsymbol{\nu\cdot\theta} }=0,\ \ \sum
p_j+\sum\nu_k \ {\rm is \ odd}}=P_{\boldsymbol{r\cdot x},
\boldsymbol{\mu\cdot\theta}}(q)|_{  \sum p_j+\sum\nu_k \ {\rm is \
even}},
\end{aligned}$$
into (3.18) then leads to the formula (3.17).
 $\square$

This theorem implies that the super binary Bell polynomials (3.12)
can still be ``linearized" by means of the Hopf-Cole transformation
$v=\ln \psi$,  $\psi=F/G$.  The formulae (3.6) and (3.17) will then
provide a way to find  the associated Lax system of supersymmetric
equations.

Finally, let's  through a graph describe  general  procedure how to
use the theory of super Bell polynomials that we have developed
above.

$$\begin{array}{ccccc}
 &   &    {\rm SUSY\ equation:}\ {\footnotesize F(\Phi)=0}        &          &
 \\[4pt]
  &   &    \vcenter{ \llap{ } }{\Big\downarrow} \vcenter{ \rlap{
  {{\footnotesize\rm dimensionless\ field:\ }q} } }    &          &      \\[4pt]
 {}   & &{\rm Bell}\ P(q)\ {\rm \ system:}
{\footnotesize E(q)=0} &
   \xrightarrow[]{q=2\ln G}   &    {{\bf\small Bilinear\ form}}    \\[4pt]
  &   &    \vcenter{ \llap{  } }{\Big\downarrow} \vcenter{ \rlap{ {{\footnotesize\rm Two-fold condition }}} } & &  \\[8pt]
  &   &   {}\ E( \widetilde{q}  )-E(q)=0       &          &          \\[4pt]
  &   &    \vcenter{ \llap{ } }{\Big\downarrow} \vcenter{ \rlap{ {{\footnotesize constaint}} } }    &          &      \\[4pt]
 {{\bf\small Binlear \ BT}}   &
\xleftarrow[w=\ln FG]{v=\ln F/G} &{\rm Binary\ Bell}\ \
{\footnotesize \mathcal{Y}(v,w)}\ {\rm \  system}     &
   \xrightarrow[]{ {{v=\ln\psi}}}   &
      {{\bf\small Lax \ pair}}    \\[4pt]
  &   &    \vcenter{ \llap{{\footnotesize }  } }{\Big\downarrow} \vcenter{ \rlap{
  {{\footnotesize $
  w=h_1(\eta,q),\ \
  v=h_2(\eta,q)$}}} }       &          &        \\[4pt]
  &    &    {{\bf\small Infinite\ conservation \ laws }}       &          &          \\[4pt]
\end{array}$$
 It is clear from this graph to see  the close connections among Bell
polynomials with bilinear equation, bilinear B\"{a}cklund
transformation, Lax pair and conservation laws.
\\[12pt]
%%%%%%%%%%%%%%%%%%%%%%%%%%%%%%%%%%%%%%%%%%%%%%%%%%%%%%%%%%%%%%%%%%%%%%%%%%%%%%%%%%%%%%%%%%%%%%%%%%%%%%%%%%%%%%%%%%%%%%%%%%%%%%%%%%%%%%
%%%%%%%%%%%%%%%%%%%%%%%%%%%%%%%%%%%%%%%%%%%%%%%%%%%%%%%%%%%%%%%%%%%%%%%%%%%%%%%%%%%%%%%%%%%%%%%%%%%%%%%%%%%%%%%%%%%%%%%%%%%%%%%%%%%%%%
{\bf\large 4. The   supersymmetric  KdV equation}\\

 Consider the supersymmetric KdV equation of Manin-Radul-Mathieu \cite{Manin,Mathieu}
$$\begin{aligned}
&\Phi_t+3\left(\Phi\mathcal{D}\Phi\right)_x+\Phi_{3x}=0,
\end{aligned}\eqno(4.1)$$
 where $\Phi=\Phi(x,t,\theta): \mathbb{R}_{\Lambda}^{2,1} \rightarrow
\Lambda_1$ is a fermionic  super field function with independent
variables $x$, $t$ and
 Grassmann variable $\theta$.  The symbol $\mathcal{D}=\partial_{\theta}+\theta\partial_x$
 denotes the super derivative differential operator, which satisfies $\mathcal{D}^2=\partial_x, \ \ \theta^2=0$.
The supersymmetric version of the KdV equation (4.1)  describe the
time evolution of  a Grassmann-valued superfield
$\Phi(x,t,\theta)=\tilde{u}(x,t)+\theta u(x,t)$, where $u(x,t)$ is
an ordinary function and $\tilde{u}(x,t)$ is a Grassmann valued
function. The variable $x, t$  acquire a Grassmann partner $\theta$,
so $(x, t, \theta)$  are coordinates in a one dimensional superspace
$\mathbb{R}_\Lambda^{2,1}$.    Since the introduction of the
supersymmetric  KdV equation (1.1) by  Manin, Radul  and Mathieu
\cite{Manin, Mathieu},   much attention has been  given to its
mathematical structure and integrable properties.  For instances,
bi-Hamiltonian structure, Painlev\'{e} property, infinite many
symmetries, Darboux transformation, B\"{a}cklund transformation,
bilinear form, super soliton solutions and super quasi-periodic
solutions  had been investigated in \cite{Oevel}--\cite{Hon}.  Here
we see how to apply the  super polynomials  to investigate complete
integrability of the supersymmetric KdV equation (4.1).

{\bf Theorem 7.}   Under the transformation $\Phi=2\mathcal{D}(\ln
G)_{x}$, the supersymmetric KdV equation (4.1) can be bilinearized
into
$$(SD_t+SD_x^3)G\cdot G=0.\eqno(4.2)$$

{\it Proof.}   The invariance of the equation (4.1) under the scale
transformation
$$x\rightarrow \lambda x, \  \ t\rightarrow \lambda^3 t, \ \ \theta\rightarrow \lambda^{{1}/{2}}\theta, \ \ \Phi\rightarrow \lambda^{-3/2}\Phi$$
shows that the dimension of the fermionic field $\Phi$ is $-3/2$,
and it can be related to a dimensionless bosonic field $q:
\mathbb{R}_{\Lambda}^{2,1} \rightarrow \Lambda_0 $,  by setting
$$\Phi=c\mathcal{D}q_{x},\eqno(4.3)$$
with $c\in \Lambda_0$ being free function to be the appropriate
choice such that the equation (4.1) connects with $P$-polynomials.
Substituting (4.3) into (4.1) and integrating with respect to $x$
yields
$$E(q)\equiv
\mathcal{D}q_{t}+\mathcal{D}q_{3x}+3cq_{2x}\mathcal{D}_{\theta}q_{x}=0.\eqno(4.4)$$
Comparing the last two terms  of this equation with the formula
(3.16) implies that we should require $c=1$. The  equation (4.4) is
then cast into a combination form of $P$-polynomials
$$E(q)=P_{t,\theta}(q)+P_{3x,\theta}(q)=0.\eqno(4.5)$$

 Making a change of dependent variable
$$q=2\ln G,  \ \ \Longleftrightarrow \ \ \Phi=2\mathcal{D}(\ln
G)_{x}$$ with $G: \mathbb{R}_{\Lambda}^{2,1} \rightarrow \Lambda_0$,
then the property (3.16) shows that the equation (4.5)  is
equivalent to  the bilinear equation (4.2). $\square$

Starting from the  bilinear equation (4.2),  it  is easy to get
super soliton solutions. For example, the regular one-soliton like
solution reads
 $$\begin{aligned}
&\Phi=\mathcal{D}[\ln(1+\exp(kx-k^3t+\theta\zeta)]_x,
\end{aligned}$$
where $k\in \Lambda_0, \zeta\in \Lambda_1$.
 Since  solving the equation (4.1) is not our main purpose in
this paper, the super soliton solutions can be found  in details
\cite{Carstea1}.

 Next,  we search for the bilinear
B\"{a}cklund transformation and Lax pair of the supersymmetric KdV
equation (4.1).

{\bf Theorem 8.}  Let $F$ be a solution of the equation (4.2), then
$G$ satisfying
$$\begin{aligned}
&(SD_x-\lambda S)F\cdot G=0,\\
&(D_t+D_x^3+3\lambda^2D_x-3\lambda D_x)F\cdot G=0
\end{aligned}\eqno(4.5)$$
is another solution of the equation (4.2).  This kind of
B\"{a}cklund transformation is exactly the same with that given by
Liu \cite{Liu3}

{\it Proof.}   Let $q=2\ln G,\ \widetilde{q}=2\ln F:
\mathbb{R}_{\Lambda}^{2,1} \rightarrow \Lambda_0 $ be  two different
solutions of the equation (4.4), respectively,  we associate the
two-field condition
 $$\begin{aligned}
&E(\widetilde{q})-E(q)=\mathcal{D}(\widetilde{q}-q)_{t}+\mathcal{D}(\widetilde{q}-q)_{3x}+3\widetilde{q}_{2x}
\mathcal{D}\widetilde{q}_x-3{q}_{2x}\mathcal{D}{q}_x=0.
\end{aligned}\eqno(4.6)$$
This two-field condition can be regarded as a  ansatz for a bilinear
B\"{a}cklund transformation and  may produce the required
transformation under appropriate additional constraints.

To find such  constraints,  we  introduce two new variables
$$v=(\widetilde{q}-q)/2=\ln F/G, \ \ w=(\widetilde{q}+q)/2=\ln FG,\eqno(4.7)$$
and   rewrite the condition (4.6) into the form
 $$\begin{aligned}
&E(\widetilde{q})-E(q)=2\mathcal{D}v_{t}+2\mathcal{D}v_{3x}+6v_{2x}\mathcal{D}w_{x}+6w_{2x}\mathcal{D}v_{x}\\
&=2\mathcal{D}[\mathcal{Y}_t(v)+\mathcal{Y}_{3x}(v,w)]+6R(v,w)=0,
\end{aligned}\eqno(4.8)$$
with
$$R(v,w)=v_{2x}\mathcal{D}w_{x}-v_{x}\mathcal{D}w_{2x}-v_x^2\mathcal{D}v_x={\rm Wronskian}[\mathcal{Y}_{x,\theta}(v,w),
\mathcal{Y}_x(v)].$$

In order to decouple  the two-field condition (4.8) into a pair of
constraints, we impose such a constraint which enable us to  express
$R(v,w)$ as the $\mathcal{D}$-derivative of a  combination of
$\mathcal{Y}$-polynomials.  A possible choice of such constraint may
be
$$\mathcal{Y}_{x,\theta}(v,w)=\lambda\mathcal{Y}_\theta(v),\eqno(4.9)$$
where  $\lambda\in \Lambda_0$  is an arbitrary parameter.  It
follows from the identity  (4.9) that
$$(v_x\mathcal{D}v)_x=\lambda\mathcal{D_{\theta}}v_x-\mathcal{D}w_{2x},$$
on account which,  then   $R(v,w)$ can be rewritten  in the form
 $$\begin{aligned}
 &R(v,w)=\lambda (v_x\mathcal{D}v)_x-2\lambda v_x\mathcal{D}v_x=\lambda^2\mathcal{D}v_x
 -\lambda\mathcal{D}w_{2x}-2\lambda v_x\mathcal{D}v_x\\
 &=\mathcal{D}[\lambda^2\mathcal{Y}_x(v)-\lambda\mathcal{Y}_{2x}(v,w)].
 \end{aligned}\eqno(4.10)$$
Then from (4.7)-(4.10),  we deduce  a coupled
 system of  super binary Bell $\mathcal{Y}$-polynomials
$$\begin{aligned}
&\mathcal{Y}_{x,\theta}(v,w)-\lambda\mathcal{Y}_\theta(v)=0,\\
&\mathcal{Y}_t(v)+\mathcal{Y}_{3x}(v,w)+
3\lambda^2\mathcal{Y}_x(v)-3\lambda\mathcal{Y}_{2x}(v,w)=0.
\end{aligned}\eqno(4.11)$$
 By application of the identity (3.15),  under transformation $v=\ln F/G, w=\ln FG$, the system (4.11)
then leads  to   the bilinear B\"{a}cklund transformation (4.5).
$\square$

 {\bf Theorem 9.}  The supersymmetric KdV equation (4.1) admits a Lax pair
$$\begin{aligned}
&(\partial_x^2+\Phi\mathcal{D}-\lambda\partial_x)\varphi=0,  \\
&[\mathcal{D}\partial_t+\mathcal{D}\partial_x^3-3\lambda\mathcal{D}\partial_x^2+3(\mathcal{D}\Phi+\lambda)\mathcal{D}
+(\mathcal{D}\Phi)\mathcal{D}]\varphi=0,
\end{aligned}\eqno(4.12)$$
where $\varphi: \mathbb{R}_{\Lambda}^{2,1} \rightarrow \Lambda_1 $
is a fermionic  eigenfunction.

{\it Proof.}  By transformation  $v=\ln \psi$, it follows  from  the
formulae (3.6) and (3.17) that
$$\begin{aligned}
&\mathcal{Y}_{\theta}(v)=\mathcal{D}\psi/\psi, \ \ \mathcal{Y}_{x,\theta}(v,w)=\mathcal{D}q_x+\mathcal{D}\psi_x/\psi,  \\
&\mathcal{Y}_{t}(v)=\psi_{t}/\psi,\
 \ \mathcal{Y}_{2x}(v,w)=q_{2x}+\psi_{2x}/\psi,\ \
\mathcal{Y}_{3x}(v,w)=3q_{2x}\psi_x/\psi+\psi_{3x}/\psi,
\end{aligned}$$
on account of which,  the system (4.12) is then linearized into a
Lax pair with a parameter  $\lambda$
$$\begin{aligned}
&L_1\psi\equiv(\mathcal{D}\partial_x-\lambda
\mathcal{D}+\mathcal{D}q_{x})\psi=0,\\
 &L_2\psi\equiv
(\partial_t+\partial_x^3+3q_{2x}\partial_x
+3\lambda^2\partial_x-3\lambda\partial^2+q_{2x})\psi=0,
\end{aligned}$$
which is  equivalent to the formula (4.12) b by  replacing
$\mathcal{D}q_{x}$ with  $\Phi$,  and  $\psi$   with
$\mathcal{D}\varphi$.
   It is easy to check that the integrability condition of the Lax
pair
$$[L_1, L_2]\psi=0$$
 is satisfied if $\Phi$ is a solution of  the
supersymmetric KdV equation (4.1). $\square$

 Finally, we  show
how to derive the infinite conservation laws for super KdV equation
(4.1) based on the use  of the binary Bell polynomials.

{\bf Theorem 10.}  The supersymmetric KdV equation (4.1) possesses
the following infinite conservation laws
$$I_{n,t}+\mathcal{D}F_{n}=0, \ n=1, 2, \cdots.\eqno(4.13)$$
where the  fermionic conserved densities $I_n's$ are  explicitly
given by recursion relations
$$\begin{aligned}
&I_1=\mathcal{D}q_{x}=\Phi,\ \ \ I_2=I_{1,x}=\Phi_x,\\
 &I_{n+1}=I_{n,x}+\sum_{k=1}^{n}I_k \mathcal{D}I_{n-k}, \ \ n=2,
3, \cdots,
\end{aligned}\eqno(4.14)$$
and the bosonic fluxes $F_n's$ are given by recursion formulas
$$\begin{aligned}
&F_1=\mathcal{D}\Phi_{2x}+3\Phi\Phi_x+3(\mathcal{D}\Phi)^2,\\
&F_2=\mathcal{D}\Phi_{3x}+3(\Phi\Phi_{2x}+\Phi_x^2)+6\mathcal{D}\Phi\mathcal{D}\Phi_x,\\
 &F_{n}=\mathcal{D}I_{n,2x}+3\sum_{k=1}^{n}(I_k I_{n+1-k}+\mathcal{D}I_k \mathcal{D}I_{n+1-k,x})+3\mathcal{D}\Phi\mathcal{D}I_n\\
 &+ \sum_{i+j+k=n}\mathcal{D}I_i\mathcal{D}I_j
 \mathcal{D}I_{k},\ \ n=3, 4, \cdots.
\end{aligned}\eqno(4.15)$$

{\it Proof. }   The conservation laws  actually have been hinted in
the two-filed constraint system (4.9)-(4.11), which  can be
rewritten in the conserved form
$$\begin{aligned}
&\mathcal{Y}_{x,\theta}(v,w)-\lambda\mathcal{Y}_\theta(v)=0,\\
&\partial_t\mathcal{Y}_{\theta}(v)+\mathcal{D}[\mathcal{Y}_{3x}(v,w)+
3\lambda^2\mathcal{Y}_x(v)-3\lambda\mathcal{Y}_{2x}(v,w)]=0.
\end{aligned}\eqno(4.16)$$
by applying the relation
$\mathcal{D}\mathcal{Y}_t(v)=\partial_t\mathcal{Y}_{\theta}(v)=\mathcal{D}v_{t}.$

 By introducing a new fermionic  potential function
  $$\eta=(\mathcal{D}\widetilde{q}-\mathcal{D}q)/2:
  \
\mathbb{R}_{\Lambda}^{2,1} \rightarrow \Lambda_1, $$ it follows from
the relation (4.8) that
$$\mathcal{D}v=\eta, \ \ \mathcal{D}w=\eta+\mathcal{D}q.\eqno(4.17)$$
Substituting (4.17) into (4.16), we get a super Riccati-type
equation
 $$\begin{aligned}
&\eta_x+\eta\mathcal{D}\eta+\mathcal{D}q_{x}-\lambda\eta=0,
\end{aligned}\eqno(4.18)$$
and a divergence-type equation
$$\begin{aligned}
&\eta_t+\mathcal{D}[\mathcal{D}\eta_{2x}+3\lambda\eta\eta_x+3q_{2x}\mathcal{D}\eta+3\mathcal{D}\eta\mathcal{D}\eta_x+(\mathcal{D}\eta)^3]=0,
\end{aligned}\eqno(4.19)$$
where we have used the equation (4.18) to get the equation (4.19).

To proceed, inserting the expansion
$$\eta=\sum_{n=1}^{\infty} I_n(\mathcal{D}q,
q_x,\cdots)\lambda^{-n},\eqno(4.20)$$
 into the equation (4.18)
and  equating the coefficients for power of $\lambda$, we then
obtain the formulas (4.13).

Finally, substituting (4.20) into (4.19) yields
$$\begin{aligned}
&\sum_{n=1}^{\infty}
I_{n,t}\lambda^{-n}+\mathcal{D}\left[\sum_{n=1}^{\infty}\mathcal{D}I_{n,2x}
\varepsilon^{-n}+3\lambda\sum_{n=1}^{\infty}I_{n}
\lambda^{-n}\sum_{n=1}^{\infty}I_{n,x}
\lambda^{-n}+3q_{2x}\sum_{n=1}^{\infty}\mathcal{D}I_{n}\lambda^{-n}
 \right.\\
&\left.+3\sum_{n=1}^{\infty}\mathcal{D}I_{n}\lambda^{-n}\sum_{n=1}^{\infty}\mathcal{D}I_{n,x}\lambda^{-n}
+(\sum_{n=1}^{\infty}\mathcal{D}I_{n}\lambda^{-n})^3\right]=0,
\end{aligned}$$
which leads to   infinite consequence of  conservation law equation
(4.13) by equating the coefficients for power of $\lambda$.
$\square$

It follows from the conservation equation (4.13) by using  (2.2)
that
$$\left(\iint I_ndxd\theta\right)_t=-\iint (\mathcal{D} F_n)dx
d\theta=0,$$ which implies that $I_n's$ are fermionic conserved
densities.  We present recursion formulas for generating an infinite
sequence of conservation laws for each equation, the first few
conserved density and associated flux  are explicit. The first
equation of conservation law equation (4.13) is exactly  the
supersymmetric KdV equation (4.1).   In conclusion, the
supersymmetric KdV (4.1) is complete integrable in the sense that it
admits bilinear B\"{a}cklund transformation, Lax pair and infinite
conservation laws.
\\[12pt]
%%%%%%%%%%%%%%%%%%%%%%%%%%%%%%%%%%%%%%%%%%%%%%%%%%%%%%%%%%%%%%%%%%%%%%%%%%%%%%%%%%%%%%%%%%%%%%%%%%%%%%%%%%%%%%%%%%%%%%%%%%%%%%%%%%%%%%
{\bf\large 5.  The  supersymmetric sine-Gordon   equation}\\

The classical sine-Gordon equation
$$\phi_{xt}=\sin\phi\eqno(5.1)$$
 has applications in various
areas of physics including nonlinear field theory, solid-state
physics, nonlinear optics, elementary particle theory and fluid
dynamics, see \cite{Lamb}-\cite{Ab1} and references therein.   The
supersymmetric extension of the equation (5.1), i.e. the
supersymmetric sine-Gordon equation \cite{Vecchia}-\cite{Grundland}
$$\begin{aligned}
&\mathcal{D}_1\mathcal{D}_2\Phi=\sin\Phi
\end{aligned}\eqno(5.2)$$ is  constructed on the four dimensional
superspace $(x, t, \theta_1, \theta_2)\in
\mathbb{R}_{\Lambda}^{2,2}$ .
  Here, $\Phi=\Phi(x, t,
\theta_1, \theta_2):\mathbb{R}_{\Lambda}^{2,2}\rightarrow \Lambda_0$
is a  scalar bosonic superfield;
  The variables  $x$ and $t$  represent the even
coordinates on the two-dimensional super-Minkowski space, while the
quantities $\theta_1$ and $\theta_2$ are anticommuting odd
coordinates which satisfy the anticommutation relations
$$\theta_1^2=\theta_2^2=0, \ \ [\theta_1, \theta_2]=0.$$
The $\mathcal{D}_1=\partial_{\theta_1}+\theta_1\partial_x$ and
$\mathcal{D}_2=\partial_{\theta_2}+\theta_2\partial_t$ are two
covariant derivatives which satisfy the anticommutation relations
$$\mathcal{D}_1^2=\partial_x, \ \ \mathcal{D}_2^2=\partial_t, \ \ [\mathcal{D}_1, \mathcal{D}_2]=0.$$

The supersymmetric version of the sine-Gordon equation was
introduced  from purely physical motivations \cite{Vecchia}. It is
becoming increasingly interesting to investigate the supersymmetric
sine-Gordon equation because of its close relation to string
theories and statistical physics \cite{ Ge}-\cite{Gr}. In recent
publications, a superspace extension of the Lagrangian formulation
has been established for the supersymmetric sine-Gordon equation
\cite{Liu4}.   The bilinear method is used to construct multi-super
soliton solutions \cite{Grammaticos}. The supersymmetric sine-Gordon
equation admits a Lax pair, and a connection was established between
its super-Backlund and super-Darboux transformations
\cite{Siddiq1,Siddiq2}.    The method of symmetry reduction is
systematically applied in order to derive invariant solutions of the
supersymmetric sine-Gordon equation \cite{Grundland}.  The
prolongation method of Wahlquist and Estabrook was used to find an
infinite-dimensional superalgebra and the associated super Lax pairs
\cite{Omote}.

 Here we  study the integrable properties of the supersymmetric
sine-Gordon based on the use of generalized super  Bell polynomials.
The bilinear form, bilinear Backlund transformation, Lax pair and
infinite conservation laws  systematically are  obtained   with our
method.

{\bf Theorem 11.}   Under  the transformation
$$\Phi=2i\ln(F/G),$$
 the  supersymmetric sine-Gordon  equation (5.2) admits
 the bilinear form
$$\begin{aligned}
&2S_1S_2F\cdot F+G^2=0,\ \ 2S_1S_2G\cdot G+F^2=0,
\end{aligned}\eqno(5.3)$$
where $F, G: \mathbb{R}_{\Lambda}^{2,2}\rightarrow \Lambda_0$ are
two bosonic functions.

 {\it Proof.}  As before, the invariance of the supersymmetric sine-Gordon  equation (5.2) under the scale
transformation
$$x\rightarrow \lambda x, \  \ t\rightarrow \lambda^{-1} t, \ \ \theta_1\rightarrow \lambda^{1/2}\theta_1, \ \ \ \ \theta_2\rightarrow \lambda^{-1/2}\theta_2,\
 \ \Phi\rightarrow \Phi$$
shows that the dimension of the bosonic superfield $\Phi$  is zero,
and so we may introduce a dimensionless bosonic field $q$ by setting
$$\Phi=cq,\eqno(5.4)$$
in which $c\in \Lambda_0$ is free constant to be determined.
Substituting (5.4) into (5.2)  yields
$$2\mathcal{D}_1\mathcal{D}_2q=P_{\theta_1\theta_2}(p+q)-P_{\theta_1\theta_2}(p-q)
=i(e^{-icq}-e^{i c q})/c\eqno(5.5)$$ where $p:
\mathbb{R}_{\Lambda}^{2,2}\rightarrow \Lambda_0$ is an auxiliary
function. If one chooses the constant $c=2i$, the equation (5.5) is
then cast into a linear combination form of $P$-polynomials
$$2P_{\theta_1\theta_2}(p+q)-2P_{\theta_1\theta_2}(p-q)+\exp(-2q)-\exp(2 q)=0,$$
which can be decoupled into a system
$$\begin{aligned}
&E_1(p,q)=2P_{\theta_1\theta_2}(p+q)+\exp(-2q)=0,\\
&E_2(p, q)=2P_{\theta_1\theta_2}(p-q)+\exp(2 q)=0.
\end{aligned}\eqno(5.6)$$
Multiplying the first equation by $\exp(p+q)$, the second equation
by $\exp(p-q)$   in the equation (5.6)  yields
$$\begin{aligned}
&2\exp(p+q) P_{\theta_1\theta_2}(p+q)+\exp(p-q)=0,\\
&2\exp(p-q)P_{\theta_1\theta_2}(p-q)+\exp(p+q)=0.
\end{aligned}\eqno(5.7)$$

By  transformation
$$q=\ln(F/G),\ \ p=\ln (FG) \Longleftrightarrow \ \  \Phi=2iq=2i\ln(F/G), \ \ p=\ln (FG)$$
and using the property (3.16), then the equation (5.7)  gives  the
bilinear form  (5.3)  for the  supersymmetric sine-Gordon  equation
(5.2). $\square$

{\bf Theorem 12.}  Let $(F,G)$ be a solution of the equation (5.3),
then $(\widetilde{F},\widetilde{G})$ satisfying
$$\begin{aligned}
&S_1\widetilde{G}\cdot G=\lambda g \widetilde{F}F, \ \
S_1\widetilde{F}\cdot F=-\lambda g \widetilde{G}G, \\
&S_2\widetilde{F}\cdot G=\frac{1}{4\lambda} g\widetilde{G}F, \ \
S_2\widetilde{G}\cdot F=-\frac{1}{4\lambda}g
\widetilde{F}G,\\
&\mathcal{D}_1g=\lambda\left(\frac{F\widetilde{F}}{G\widetilde{G}}-\frac{G\widetilde{G}}{F\widetilde{F}}\right),\
\
\mathcal{D}_2g=\frac{1}{4\lambda}\left(\frac{G\widetilde{F}}{F\widetilde{G}}-\frac{F\widetilde{G}}{G\widetilde{F}}\right).
\end{aligned}\eqno(5.8)$$
is another solution of the equation (5.3), where
$g:\mathbb{R}_{\Lambda}^{2,2}\rightarrow \Lambda_1 $ fermionic
auxiliary superfield  and  $\lambda\in \Lambda_0$ is even parameter.

{\it Proof.}
 In
order to obtain the bilinear B\"{a}cklund transformation and Lax
pairs of the equation (5.2),  let $p, q$ and $\widetilde{p},
\widetilde{q}$  be  two  solutions of the equation (5.6) and
consider the associated two-field condition
 $$\begin{aligned}
&E_1(\widetilde{p},\widetilde{q} )-E_1(p,
q)=2\mathcal{D}_1\mathcal{D}_2(\widetilde{p}-p)-2\mathcal{D}_1\mathcal{D}_2
(\widetilde{q}-q)\\
&\ \ \ \ \ \
 \ \ \ \  \ \ \ \ \
 \ \ \ \  \ \ \ \ \
 \ \ \ \ +e^{\widetilde{q}+q}(e^{\widetilde{q}-q}-e^{q-\widetilde{q}})=0,\\
&E_2(\widetilde{p},\widetilde{q} )-E_2(p,
q)=2\mathcal{D}_1\mathcal{D}_2(\widetilde{p}-p)+2\mathcal{D}_1\mathcal{D}_2
(\widetilde{q}-q)\\
&\ \ \ \ \ \
 \ \ \ \  \ \ \ \ \
 \ \ \ \  \ \ \ \ \
 \ \ \ \ +e^{-(\widetilde{q}+q)}(e^{\widetilde{q}-q}-e^{q-\widetilde{q}})=0,
\end{aligned}\eqno(5.9)$$
 where
$$\begin{aligned}
&\widetilde{p}=\ln(\widetilde{F}\widetilde{G}),\ \
\widetilde{q}=\ln(\widetilde{F}/\widetilde{G}),
\end{aligned}$$
We  introduce variables
$$\begin{aligned}
&v_1=\ln(\widetilde{G}/G), \ v_2=\ln(\widetilde{F}/F),\
v_3=\ln(\widetilde{F}/G),\  v_4=\ln(\widetilde{G}/F),\\
&w_1=\ln(\widetilde{G}G),\  w_2=\ln(\widetilde{F}F),\
w_3=\ln(\widetilde{F}G),\ w_4=\ln(\widetilde{G}F),
\end{aligned}$$
from which, we have relations
$$\begin{aligned}
&\widetilde{q}-q=v_2-v_1=w_3-w_4, \ \
\widetilde{q}+q=v_3-v_4=w_2-w_1,\\
&\widetilde{p}-p=v_1+v_2=v_3+v_4, \ \
\widetilde{p}+p=w_1+w_2=w_3+w_4
\end{aligned}\eqno(5.10)$$
and
$$\begin{aligned}
&v_1=v_4+q,\ \ v_2=v_3-q, \ \ w_1=w_4-q, \ \ w_2=w_3+q.
\end{aligned}\eqno(5.11)$$

By using the mixed variables (5.10),  it follows that from (5.9)
 $$\begin{aligned}
&4\mathcal{D}_1\mathcal{D}_2
v_1+e^{v_3-v_4}(e^{v_2-v_1}-e^{v_1-v_2})=0,\\
&4\mathcal{D}_1\mathcal{D}_2
v_2+e^{v_4-v_3}(e^{v_1-v_2}-e^{v_2-v_1})=0,
\end{aligned}\eqno(5.12)$$
which may produce the required bilinear B\"{a}cklund transformation
under an appropriate additional constraint.  We choose a constraint
$$\mathcal{D}_1v_1=\mathcal{Y}_{\theta_1}(v_1)=\lambda g e^{v_3-v_4},\eqno(5.13)$$
where $g:\mathbb{R}_{\Lambda}^{2,2}\rightarrow \Lambda_1 $ fermionic
auxiliary superfield  and  $\lambda\in \Lambda_0$ is even parameter.
The fermionic function $g$ is introduced because of supersymmetry
and the oddness of the superspace derivatives $\mathcal{D}_1,
\mathcal{D}_2$.  The constraint (5.13)  reduces the first equation
in (5.12) into
$$-4\lambda\mathcal{D}_2g-4\lambda g\mathcal{D}_2(v_3-v_4)+e^{v_2-v_1}-e^{v_1-v_2}=0.\eqno(5.14)$$
Since the term  $\mathcal{D}_2(v_3-v_4)$ should be fermionic
function, we make a constraint
$$\mathcal{D}_2(v_3-v_4)=\frac{1}{4\lambda} g h,\eqno(5.15)$$
where $h$ is a bosonic function to be determined. On account of this
constraint, it follows from (5.14) that
$$\mathcal{D}_2g=\frac{1}{4\lambda}(e^{v_2-v_1}-e^{v_1-v_2}),\eqno(5.16)$$
which holds because $g^2=0$, $g$ being fermionic.

By means of the system (5.15) and (5.16), the second equation in
(5.12) reads
$$\begin{aligned}
&\mathcal{D}_2 (\mathcal{D}_1v_2+\lambda ge^{v_4-v_3})=0,
\end{aligned}$$
which is satisfied if we choose
$$\begin{aligned}
&\mathcal{Y}_{\theta_1}(v_2)=\mathcal{D}_1v_2=-\lambda ge^{v_4-v_3}.
\end{aligned}\eqno(5.17)$$

On the one hand, using the relation (5.11),  we have
$$\begin{aligned}
&4\mathcal{D}_1 \mathcal{D}_2(v_3-v_4)+4\mathcal{D}_2
\mathcal{D}_1(v_1-v_2) =8\mathcal{D}_1
\mathcal{D}_2q\\
&=(e^{v_2-v_1}-e^{v_1-v_2})(e^{v_3-v_4}+e^{v_4-v_3})+(e^{v_2-v_1}+e^{v_1-v_2})(e^{v_3-v_4}-e^{v_4-v_3}).
\end{aligned}\eqno(5.18)$$
On the other hand, it follows from (5.13), (5.16) and (5.17) that
$$\begin{aligned}
&4\mathcal{D}_2 \mathcal{D}_1(v_1-v_2)
=4\lambda(\mathcal{D}_2g)(e^{v_3-v_4}+e^{v_4-v_3}) \\
&=(e^{v_2-v_1}-e^{v_1-v_2})(e^{v_3-v_4}+e^{v_4-v_3}).
\end{aligned}\eqno(5.19)$$
Combining (5.15), (5.18) and (5.19) yields
$$\begin{aligned}
&4\mathcal{D}_2 \mathcal{D}_1(v_3-v_4)
=\frac{1}{\lambda}(\mathcal{D}_1g)h=(e^{v_3-v_4}-e^{v_4-v_3})(e^{v_2-v_1}+e^{v_1-v_2}),
\end{aligned}$$
which implies that we may choose
$$\begin{aligned}
&\mathcal{D}_1g=\lambda(e^{v_3-v_4}-e^{v_4-v_3})
\end{aligned}\eqno(5.20)$$
and
$$\begin{aligned}
& h =(e^{v_2-v_1}+e^{v_1-v_2}).
\end{aligned}$$
 Thus, we have
$$\begin{aligned}
&\mathcal{D}_2(v_3-v_4)
=\frac{1}{4\lambda}g(e^{v_2-v_1}+e^{v_1-v_2}),
\end{aligned}$$
which can be written as a pair of $\mathcal{Y}$-polynomials
$$\mathcal{Y}_{\theta_2}(v_3)=\mathcal{D}_2v_3=\frac{1}{4\lambda}ge^{v_1-v_2},\ \ \mathcal{Y}_{\theta_2}(v_4)
=\mathcal{D}_2v_4 =-\frac{1}{4\lambda}ge^{v_2-v_1},\eqno(5.21)$$

Combining (5.13), (5.16), (5.17), (5.20) and (5.21)  gives bilinear
B\"{a}cklund transformation (5.8) of the supersymmetric sine-Gordon
equation. $\square$

Finally we derive Lax pair of the supersymmetric sine-Gordon
equation.

 {\bf Theorem 13.} The supersymmetric sine-Gordon equation
(5.2) admits a Lax pair
$$\begin{aligned}
&\mathcal{D}_1\Psi=M\Psi=\left(
\begin{matrix}\displaystyle{-\frac{1}{2}}i\mathcal{D}_1\Phi&\lambda g\cr
\displaystyle{\lambda g}&\displaystyle{\frac{1}{2}}i\mathcal{D}_1\Phi\end{matrix}\right)\Psi,\\[4pt]
&\mathcal{D}_2\Psi=N\Psi=\left(
\begin{matrix}0&\displaystyle{-\frac{1}{4\lambda }}ge^{-i\Phi}\cr
\displaystyle{-\frac{1}{4\lambda
}ge^{i\Phi}}&0\end{matrix}\right)\Psi,
\end{aligned}\eqno(5.22)$$
together with
$$\begin{aligned}
&\mathcal{D}_1g=\lambda\left(\frac{\psi_3}{\psi_4}-\frac{\psi_4}{\psi_3}\right),\
\
\mathcal{D}_2g=\frac{1}{\lambda}\left(e^{i\Phi}\frac{\psi_3}{\psi_4}-e^{-i\Phi}\frac{\psi_4}{\psi_3}\right),
\end{aligned}$$
where $\Psi=(\psi_3, \psi_4)^T$.

 Making use of the Hopf-Cole  transformation
 $$v_3=\ln \psi_3, \ \ v_4=\ln \psi_4,$$
  then the system (5.13), (5.16),
(5.17), (5.20) and (5.21) can be linearized into a Lax pair (5.22).
It is easy to check that the integrability condition
$$\mathcal{D}_2M+\mathcal{D}_1N-[M, N]=0$$
is satisfied if $\Phi$ is a solution of  the sine-Gordon equation
(5.2).

If we choose a transformation
$$ \phi_1=\psi_4^2, \ \ \phi_2=\psi_3^3,  \ \ g=\frac{\phi_3}{2i
\psi_3\psi_4}$$
 then the Lax pair (5.22) is also equivalent to  a linear system  in $3\times 3$
matrix form
$$\begin{aligned}
&\mathcal{D}_1\Omega=\frac{1}{4}\left(
\begin{matrix}4\mathcal{D}_1\Phi&0&\lambda\cr 0&-4\mathcal{D}_1\Phi&-\lambda\cr
-4\lambda&4\lambda &0\end{matrix}\right)\Omega, \\
&\mathcal{D}_2\Omega=\frac{1}{16\lambda}\left(
\begin{matrix}0&0&e^{i\Phi}\cr 0&0&-e^{-i\Phi}\cr
4e^{-i\Phi}&-4e^{-i\Phi}&0\end{matrix}\right)\Omega,
\end{aligned}\eqno(5.23)$$
where $\Omega=(\phi_1, \phi_2, \phi_3)^T$, $\phi_1,
\phi_2:\mathbb{R}_{\Lambda}^{2,2}\rightarrow  \Lambda_0$ are bosonic
functions and $\phi_3:\mathbb{R}_{\Lambda}^{2,2}\rightarrow
\Lambda_1$ is a fermionic function. The system (5.23) also can be
obtained  from (5.13), (5.16), (5.17), (5.20) and (5.21) by setting
$$2(v_4-v_3)=\ln\frac{\phi_1}{\phi_2}, \ \
2(v_2-v_1)=-2i\Phi+\ln\frac{\phi_1}{\phi_2}, \ \
g=\frac{\phi_3}{2i\sqrt{\phi_1\phi_2}}.$$ The compatibility of the
linear system (5.23) in superspace is equivalent to the equation
(5.2). The system (5.23) is the same as obtained in \cite{Sciuto},
but here it is derived systematically from the super Bell
polynomials and Lax pairs. $\square$

 Noting the transformation relation
 $$v_1-v_2=i(\Phi-\widetilde{\Phi})/2,
\ \ v_3-v_4=-i(\Phi+\widetilde{\Phi})/2,$$
 then it follows  from equations  (5.13), (5.17),
(5.18), (5.21) and (5.22) that
$$\begin{aligned}
&\mathcal{D}_1(\Phi-\widetilde{\Phi})=\lambda
g\cos\left(\frac{\Phi+\widetilde{\Phi}}{2}\right), \\
&\mathcal{D}_2(\Phi+\widetilde{\Phi})=\frac{1}{4\lambda}
g\cos\left(\frac{\Phi-\widetilde{\Phi}}{2}\right),\\
&\mathcal{D}_1g=\lambda
\sin\left(\frac{\Phi+\widetilde{\Phi}}{2}\right),\ \
\mathcal{D}_2g=\frac{1}{4\lambda}
\sin\left(\frac{\Phi-\widetilde{\Phi}}{2}\right),
\end{aligned}\eqno(5.24)$$
which is the B\"{a}cklund transformation of the supersymmetric
sine-Gordon equation.
 The compatibility of
the B\"{a}cklund transformation (5.24) is the supersymmetric
sine-Gordon equation for both $\Phi$ and $\widetilde{\Phi}$
separately. The super B\"{a}cklund transformation (5.24) reduces to
the classical B\"{a}cklund transformation of the purely bosonic
sine-Gordon equation when fermions are equal to zero.
\\[12pt]
%%%%%%%%%%%%%%%%%%%%%%%%%%%%%%%%%%%%%%%%%%%%%%%%%%%%%%%%%%%%%%%%%%%%%%%%%%%%%%%%%%%%%%%%%%%%%%%%%%%%%%%%%%%%%%%%%%%%%%%%%%%%%%%%%%%%%%
%%%%%%%%%%%%%%%%%%%%%%%%%%%%%%%%%%%%%%%%%%%%%%%%%%%%%%%%%%%%%%%%%%%%%%%%%%%%%%%%%%%%%%%%%%%%%%%%%%%%%%%%%%%%%%%%%%%%%%%%%%%%%%%%%%%%%%
{\bf\large  6.  Concluding Remarks  }\\

In this paper, we have introduced a class of super Bell polynomials
which play an important role in the characterization of bilinear
B\"{a}cklund transformation, Lax pairs and infinite conservation
laws of supersymmetric equations. To the knowledge of the authors,
this is the first work on the super Bell polynomials and their
applications to super integrable systems. We believe that there are
still many interesting deep relations between generalized Bell
polynomials and integrable structures, which remain open and worth
to be considered.  For instance,  (i) How to explore the relations
between the super Bell polynomials with symmetries, Hamiltonian
functions, etc. (ii) How to define a class of discrete Bell
polynomials and apply them in discrete equations. We have some ideas
on these questions and will intend to return to
them in some future   publications.\\[12pt]
  %%%%%%%%%%%%%%%%%%%%%%%%%%%%%%%%%%%%%%%%%%%%%%%%%%%%%%%%%%%%%%%%%%%%%%%%%%%%%%%%%%%%%%%%%%%%%%%%%%%%%%%%%%%%%%%%%%%%%%%
%%%%%%%%%%%%%%%%%%%%%%%%%%%%%%%%%%%%%%%%%%%%%%%%%%%%%%%%%%%%%%%%%%%%%%%%%%%%%%%%%%%%%%%%%%%%%%%%%%%%%%%%%%%%%%%%%%%%%%
{\bf\large  Acknowledgment}

The work described in this paper was supported by grants from
  the CityU (Project No. 7002440),  the National Science Foundation of China (No.
10971031) and Shanghai Shuguang Tracking Project (No. 08GG01).


\begin{thebibliography}{99} %% The number "99" means that this list has more than nine items.

\small
 \baselineskip=23pt

\bibitem{Ramond}   P. Ramond, Dual theory for free fermions,  Phys. Rev. D  {\bf 3},  2415-2418 (1971).

\bibitem{Neveu}  A. Neveu and J. H. Schwarz, Factorizable dual model of
pions, Nucl. Phys. B  {\bf 31}, 86-112 (1971).

\bibitem{Wess}  J. Wess and B. Zumino, Supergauge transformations in four dimensions,
    Nucl. Phys. B {\bf 70},  39-50 (1974).



\bibitem{Manin}     Yu. I. Manin and A. O. Radul, A supersymmetric extension of the Kadomtsev-Petviashvili hierarchy,
Commun. Math. Phys.   {\bf 98},  65-77 (1985).

 \bibitem{} J. M. Rabin, The Geometry of the Super KP Flows, Commun. Math.
Phys. {\bf 137}, 533-552 (1991)


 \bibitem{Mathieu}   P. Mathieu,  Supersymmetric extension of the Korteweg-de Vries equation,
  J. Math. Phys.   {\bf 29},  2499-2506 (1988).

 \bibitem{}   B. A. Kupershmidt,  A super Korteweg-de Vries equation: An integrable
 system, Phys.  Lett.  A,  {\bf 102}, 213-215 ( 1984)

 Super Korteweg-de Vries equations associated to super extensions of the Virasoro algebra

 \bibitem{} B. A. Kupershmidt,  This article is not included in your
organization's subscription. However, you may be able to access this
article under your organization's agreement with Elsevier,  Phys.
Lett.  A,  {\bf 109}, 417-423  ( 1985)


  \bibitem{}  J. Harnad and B. A. Kupershmidt, Super Loop Groups, Hamiltonian Actions
and Super Virasoro Algebras, Commun. Math. Phys. 132, 315-347 (1990)

 \bibitem{Oevel}    W. Oevel and  Z. Popowicz, The bi-Hamiltonian structure of fully supersymmetric Korteweg-de
 Vries systems,  Commun. Math. Phys.  {\bf 139},  441-460 (1991).

\bibitem{Morosi}    C. Morosi, L. Pizzocchero, On the BiHamiltonian Structure
of the Supersymmetric KdV Hierarchies. A Lie Superalgebraic
Approach, Commun. Math. Phys. {\bf 158}, 267-288 (1993)

\bibitem{Inami}  T. Inami and H. Kanno, Lie superalgebraic approach to super
Toda lattice and generalized super KdV equations, Commun. Math.
Phys. {\bf 136}, 519-542 (1991)


\bibitem{} E. Ivanov, S. Krivonos, New integrable extensions of $\mathcal{N} = 2$ KdV and
Boussinesq hierarchies, Physics Letters A  {\bf 231}, 75-81  ( 1997)


 \bibitem{}  E. Ivanov,  S. Krivonos, F. Toppan,  $\mathcal{N}=4$ super NLS-mKdV hierarchies, Physics Letters B {\bf 405},
 85-94  (1997)

   \bibitem{Sarma}    D. Sarma, Constructing a supersymmetric integrable system from the Hirota method in superspace,
 Nucl. Phys. B  {\bf 681},  351-358 (2004).

 \bibitem{Liu1}   Q. P. Liu, Darboux transformations for supersymmetric korteweg-de vries equations,
    Lett Math. Phys.   {\bf 35},  115-122 (1995).

 \bibitem{Liu3}  Q. P. Liu and  X. B. Hu, Bilinearization of $\mathcal{N}=1$ supersymmetric Korteweg de Vries equation revisited,
  J.  Phys. A   {\bf 38},  371-6378 (2005).

\bibitem{Liu4}    Q. P. Liu, X. B. Hu and M. X. Zhang, Supersymmetric modified Korteweg-de Vries
       equation: bilinear approach,  Nonlinearity   {\bf 18},  1597-1603 (2005).


 \bibitem{Yung}  I. N. McArthur, C. M. Yung,  Hirota bilinear form for the super-KdV
 hierarchy,  Mod. Phys. Lett. A  18, 1739-1745 (1993)


 \bibitem{Carstea1}   A. S. Carstea, Extension of the bilinear formalism to supersymmetric KdV-type equationsnipne,
   Nonlinearity   {\bf 13},  1645-1656 (2000).

 \bibitem{Hon}     E. G. Fan and Y. C. Hon, Quasi-periodic wave solutions of  $\mathcal{N}=2$
supersymmetric KdV equation in superspace, Stud. Appl. Math. in
Press



\bibitem{Bell}   E. T. Bell, Exponential polynomials, Ann. Math.  {\bf
35}, 258-277  (1934)

\bibitem{Abr}  M. Abramowitz and J. A. Stegun, Eds., {\it Handbook of
Mathematical Functions with Formulas, Graphs and Mathematical
Tables}, Dover, New York, 1972.

\bibitem{} L. Comtet, {\it  Advanced Combinatorics}, Reidel, Dordrecht,
1974

\bibitem{Com}  J. Riordan, {\it  Combinatorial Identities}, Wiley, New York, 1966.


\bibitem{Kolbig} F. T. Howard, A special class of Bell polynomials,  Math. Comput.  {\bf 35}  977-989 (1980)


\bibitem{} S. Noschese and P.E. Ricci, Differentiation of multivariable
composite functions and Bell polynomials,  J. Comput.  Anal. Appl.
{\bf  5}, 333-340 (2003).

\bibitem{} P. Natalini and P.E. Ricci, An extension of the Bell polynomials,
Computers  Math. Appl. {\bf 47 }, 719-725 (2004).

\bibitem{} A. Bernardini, P. Natalini and P. E. Ricci, Multidimensional Bell Polynomials of Higher Order, Comput.
Math. Appl.  {\bf 50} 1697-1708 (2005)

\bibitem{Wang} R.B. Paris, The asymptotics of the generalised Hermite-Bell polynomials, J. Comput.  Appl.
Math.   {\bf 232}  216-226 (2009)


\bibitem{Gilson} C. Gilson, F. Lambert, J. Nimmo and R. Willox, On the combinatorics of the Hirota D-operators,
Proc. R. Soc. Lond. A {\bf 452}, 223-234 (1996)

\bibitem{Lambert1}  F. Lambert, I. Loris and J.Springael, Classical Darboux transformations and the KP
hierarchy, Inverse Probl. {\bf 17}, 1067-1074 (2001)

\bibitem{Lambert2}  F. Lambert and J. Springael, Soliton equations and simple combinatorics, Acta Appl.  Math.  {\bf 102}, 147-178 (2008)


\bibitem{Vlad1}    V. S. Vladimirov, Superanalysis. I. Differential
calculus,  Theor. Math. Phys.   {\bf 59},  317-335 (1984).

\bibitem{}    V. S. Vladimirov, Superanalysis. II. Integral calculus,  Theor. Math. Phys.   {\bf 60}, 743-765  (1984)

\bibitem{Berezin}  F. A. Berezin: {\it Introduction to Superanalysis}, Reidel Publishing Company,
Dordrecht, 1987.



\bibitem{Vlad2}    A. Khrennikov,  {\it Superanalysis},  Kluwer Academic Publishers, Dordrecht, 1999


\bibitem{Lamb}   G L Lamb,  {\it  Elements of Soliton Theory}, New York: Wiley, 1980


\bibitem{}   R. K. Bullough,  P. J. Caudrey and  H. M. Gibbs,  The double
sine-Gordon equations: a physically applicable system of equations
Solitons,  eds.  R. K. Bullough and P. J. Caudrey, New York,
Springer, 1980



\bibitem{}  C. Rogers  and   W. K. Schief,  {\it  Backlund and Darboux
Transformations}, Cambridge,  Cambridge University Press, 2002

\bibitem{Sciuto}      S. Sciuto, Exterior Calculus and two-dimensional supersymmetric models,
Phys.  Lett.  B   {\bf 90 },  75-80 ( 1980)

\bibitem{Ab1}  M. J. Ablowitz and  P. A. Clarkson,  Solitons, Nonlinear Evolution
Equations and Inverse Scattering, Cambridge: Cambridge University
Press, 1991

\bibitem{Vecchia}  P. Di Vecchia and S. Ferrara, Classical solutions in two-dimensional supersymmetric field theories,  Nucl. Phys. B {\bf
130},  93-104 (1977)

\bibitem{Ge}   D. Gepner,  Space-time supersymmetry in compactified string theory and superconformal models,   Nucl. Phys. B {\bf 296}, 757-778 (1988)

\bibitem{Ba}       Z Bajnok, C. Dunning, L Palla,  G. Takacs and  F. Wagner, SUSY sine-Gordon theory as a perturbed
conformal field theory and finite size effects,  Nucl. Phys. B {\bf
679}, 521-544  (2004)


\bibitem{Gr}   M. Grigoriev and  A.  A. Tseytlin, Pohlmeyer reduction of AdS(5) x S(5) superstring sigma model,
  Nucl. Phys. B {\bf 800}, 4450-501  (2008)



\bibitem{Grammaticos} B. Grammaticos, A. Ramani and A. S. Carstea, Bilinearization and soliton solutions of the $\mathcal{N} =
1$ supersymmetric sine-ordon equation, J. Phys. A, {\bf 34},
4881-4886 (2001)

\bibitem{Siddiq1} M. Siddiq  and   M. Hassan, On the linearization of the super sine-Gordon equation, Europhys. Lett. {\bf 70},
149-154 (2005)


\bibitem{Siddiq2}   M. Siddiq, M. Hassan and U Saleem, On Darboux transformation of the supersymmetric
sine-Gordon equation, J. Phys. A,  {\bf 39}, 7313-7318 (2006)



\bibitem{Grundland} A. M. Grundland, A. J. Hariton and L. Snobl,  Invariant solutions of
the supersymmetric sine-Gordon equation, J. Phys. A,  {\bf 42}
335203 (2009)

 \bibitem{Omote}    M.  Omote, Prolongation
structures of the supersymmetric Sine-Gordon equation and
infinite-dimensional superalgebras,   J. Phys. A, {\bf 20},
1941-1950 (1987)



\end{thebibliography}
\end{document}